\documentclass[11pt]{article}

\usepackage[utf8]{inputenc}
\usepackage[T1]{fontenc}
\usepackage{amsmath, amssymb, amsthm}
\usepackage{graphicx}
\usepackage{natbib}           
\usepackage{hyperref}         
\usepackage{geometry}         
\usepackage{setspace}         
\usepackage{authblk}          
\usepackage{booktabs}         
\usepackage{float}            
\usepackage{enumitem}         
\usepackage{bm}
\usepackage{comment}
\usepackage{epstopdf}
\usepackage{algorithm}
\usepackage{algpseudocode}
\usepackage{dsfont}
\usepackage{nccmath}
\newtheorem{theorem}{Theorem}
\newtheorem{lemma}[theorem]{Lemma}  

\title{\textbf{Adaptive Design for Contour Estimation from Computer Experiments with Quantitative and Qualitative Inputs}}
\author[1]{Anita Shahrokhian\thanks{
    \href{mailto:a.shahrokhian@queensu.ca}{a.shahrokhian@queensu.ca} }\hspace{.2cm}}
\affil[1]{Department of Mathematics and Statistics, Queen’s University, Canada}
\author[2] {Xinwei Deng\thanks{
    \href{mailto:xdeng@vt.edu}{xdeng@vt.edu} }\hspace{.2cm}}
\affil[2]{Department of Statistics, Virginia Tech, USA}
\author[1]{C. Devon Lin\thanks{
  \href{mailto:devon.lin@queensu.ca}{devon.lin@queensu.ca}}\hspace{.2cm}} 
\author[3]{ Pritam Ranjan\thanks{
  \href{mailto:pritamr@iimidr.ac.in}{pritamr@iimidr.ac.in}}\hspace{.2cm}} 
\affil[3]{Operations Management Quantitative Techniques, Indian Institute of Management Indore, MP, India}
\author[2]{ Li Xu\thanks{
  \href{mailto:lix1992@vt.edu} {lix1992@vt.edu})}\hspace{.2cm}}

\date{ }

\begin{document}

\maketitle

\begin{abstract}
 Computer experiments with quantitative and qualitative inputs are widely used to study many scientific and engineering processes. Much of the existing work has focused on design and modeling or process optimization for such experiments. This paper proposes an adaptive design approach for estimating a contour from computer experiments with quantitative and qualitative inputs. A new criterion is introduced to search for the follow-up inputs. The key features of the proposed criterion are (a) the criterion yields adaptive search regions; and (b) it is region-based cooperative in that for each stage of the sequential procedure, the candidate points  in the design space is divided into two disjoint groups using confidence bounds, and within each group, an acquisition function is used to select a candidate point. Among the two selected points, a point that is closer to the contour level with the higher uncertainty or that has higher uncertainty when the distance between its prediction and the contour level is within a threshold  is chosen.  The proposed approach provides empirically more accurate contour estimation than existing approaches as illustrated in numerical examples and a real application. Theoretical justification of the proposed adaptive search region is given. 
\end{abstract}

\noindent\textbf{Keywords:} Active learning, Emulator, Expected improvement, Gaussian process, Inverse problem, Sequential design, Uncertainty quantification, Upper confidence bound.


\section{Introduction}
Computer simulators (models)  have become fundamental and popular for studying complex physical systems \cite{sacks1989design,santner2003design}. They are the
mathematical models implemented in computer code. With the computer simulator at hands, computer experiments are conducted by varying the inputs and observing outputs.  The scientific values of computer experiments have been demonstrated successfully in numerous  applications such as fire evolution simulator, prosthesis device
design, rocket booster simulation, and tidal power potential prediction \cite{cooper1985aset,chang2001decolorization,gramacy2008bayesian,macdonald2015gpfit}. In some applications, computer experiments are computationally expensive as it may take several hours or days to get a single output. One common approach to address such an issue is to build statistical emulators (surrogate) to approximate computer simulators so that  emulators can be used for further statistical analysis such as optimization \cite{jones1998efficient}, contour estimation \cite{ranjan2008sequential,bichon2008efficient}, calibration \cite{kennedy2001bayesian},  sensitivity analysis \cite{oakley2004probabilistic}, and so on. 
Computer experiments with quantitative and qualitative inputs occur frequently in applications in science and engineering. For instance, \cite{rawlinson2006retrieval} studied the design of a knee prosthesis with qualitative inputs on prosthesis design and force pattern.   \cite{qian2008gaussian} considered a computer simulator for determining several air characteristics for a data storage area in which the input location of an air diffuser unit, the location of 
the hot air return vent, and the type of power unit are qualitative. For such computer experiments, various modeling approaches have been proposed to take into account the effects of qualitative inputs. The most recent efforts are the latent variable based Gaussian process models \cite{zhang2020latent} and the easy-to-interpret Gaussian process models \cite{xiao2021ezgp}. To choose inputs for performing computer experiments with quantitative and qualitative inputs, sliced Latin hypercubes, and marginally (or doubly) coupled designs have been introduced \cite{qian2012sliced, he2017marginally,he2019construction,yang2022doubly}. Although there is abundant work for the design and modeling of computer experiments with both types of inputs, little has been done for the analysis of such computer experiments. To the best of our knowledge, the only exceptions are for the problem of global optimization \cite{zhang2020bayesian,cai2024adaptive}. In this article, we consider the problem of contour (also known as level or excursion set) estimation for computer experiments with quantitative and qualitative inputs. The goal is to search for the inputs that yield a contour level of interest. 

The designs for computer experiments can be space-filling designs or model-dependent designs. Space-filling designs are chosen such that the design points are spread out the design space as uniformly as possible, regardless of the emulators chosen \cite{lin2015latin}. Thus, they are model independent. Model-dependent designs can be one-shot or adaptive designs. Space-filling or one-shot model-dependent designs can be inefficient because they do not take the goal of the analysis into account. Using adaptive designs, also known as active learning, is a practical way to overcome this issue by judiciously choosing inputs that help achieve the goal of the analysis with fewer design points. Adaptive designs use an initial design and fit the model for the data at hands and repeat the process of input selection and updating the model after each selection until they meet some stopping criterion. In adaptive designs, the essential issue is how to choose the follow-up inputs based on the current data collected.  In the literature,  various criteria such as expected improvements (EI), upper confidence bound, stepwise uncertainty
reduction, entropy search and knowledge gradient were introduced for different goals of the analysis \cite{jones1998efficient,ranjan2008sequential,picheny2010adaptive,Srinivas_2012,henrandez2014predictive,frazier2009knowledge,bect2012sequential,chevalier2014fast,marques2018contour,cole2023entropy}.

When the goal of the analysis is contour estimation, the issue of how to choose inputs adaptively for computer experiments with both quantitative and qualitative inputs has not been addressed in the literature. 
 Although existing criteria for contour estimation from computer experiments with only quantitative inputs can be used for those with mixed inputs, their performance may be poor as we will demonstrate in the numerical examples. We propose an adaptive design using confidence bounds in the contour estimation problem by grouping the candidate points and using the proper criterion for each subgroup to solve the problem for any contour level of interest. The proposed method selects the candidate points in each subgroup using a certain criterion, and between the two selected points, we choose the next candidate point such that its prediction is closer to the contour level of interest and has the larger predictive variance, or it has a larger predictive variance when the difference between its prediction and the contour level is smaller than some threshold. 

The rest of the paper is organized as follows. Section 2 provides the necessary background on Gaussian processes (GP) models for computer experiments with both types of inputs, and reviews different acquisition functions used in models with only quantitative inputs.  Section 3 presents the proposed adaptive design approach for contour estimation for computer experiments with both types of inputs. The theoretical justifications of adaptive search regions are also given. We illustrate the efficiency of the proposed method and compare it with existing methods through the simulation study in Section 4. Section 5 applies the proposed approach in a real application in high performance computing. Section 6 concludes the paper. Details of the proof are provided in the Appendix.  

\section{Background and Review}\label{sec:back}
\subsection{GP Models with Quantitative and Qualitative Inputs}
Consider an $n$-run computer experiment with $p$ quantitative variables and $q$ qualitative factors. Let $x^{(k)}$ be the $k$th quantitative variable and $z^{(h)}$ be the $h$th qualitative factor having $m_h$ levels.  For $i=1,\ldots, n$, denote $\bm{x}_i = (x_{i1},\ldots, x_{ip})^T$ and 
$\bm{z}_i = (z_{i1},\ldots, z_{iq})^T$, and let $\bm{w}_i = (\bm{x}_i^T,\bm{z}_i^T)^T$ be the $i$th input and $y_i$ be the corresponding output.  
To model the relationship between output $Y$ and an input $\bm{w} = (\bm{x}^T,\bm{z}^T)^T$,  \cite{xiao2021ezgp} proposed the model, 
\begin{align}\label{ADGP}
    Y(\bm{x},\bm{z})=\mu+G_0(\bm{x})+G_{{z}^{(1)}}(\bm{x})+\dots+G_{{z}^{(q)}}(\bm{x}),
\end{align}
where $G_0$ and $G_{{z}^{(h)}} $'s are independent GPs with mean zero and covariance function $\phi_0$ and $\phi_h$, $h=1,\dots,q$.  \cite{xiao2021ezgp} assumes that  $G_0$ is a standard GP with only quantitative inputs $\bm{x}$ with $\phi_0$ given by, 
 \begin{align}\label{EzGpcov1}
     \phi_0(\bm{x}_i,\bm{x}_j|\bm{\theta}_0)=\sigma_0^2\exp\{-\sum_{k=1}^{p}\theta_k^{(0)}(x_{ik}-x_{jk})^2\},
 \end{align}
\noindent with $\theta_k^{(0)}>0$. For $h=1, \dots,q$, the covariance function of $G_{{z}^{(h)}}$ is assumed to be,
 \begin{align}\label{EzGpcov2}
 \begin{split}
     \phi_h((\bm{x}_i^T,z_{ih})^T,(\bm{x}_j^T,z_{jh})^T|\bm{\Theta}^{(h)})&=\sigma_h^2\exp\{-\sum_{k=1}^{p}\theta^{(h)}_{kl_h}(x_{ik}-x_{jk})^2\}\mathds{1}(z_{ih}=z_{jh}\equiv l_h),\\
 \end{split}
 \end{align}
where $z_{ih}$ and $z_{jh}$ are the $i$th and $j$th value of the input variable ${z}^{(h)}$,  $\sigma_h^2$ is the variance parameter for ${z}^{(h)}$, $l_h=1,\dots,m_h$, $m_h$ is the number of the levels of the qualitative input variable ${z}^{(h)}$, $\bm{\Theta}^{(h)}=(\theta^{(h)}_{kl_h})_{p\times m_h}$ is the matrix for the correlation parameters, and the indicator function is 1 when $z_{ih}=z_{jh}\equiv l_h$ and 0 otherwise. Therefore, from (\ref{EzGpcov1}) and (\ref{EzGpcov2}), the covariance function in the model in (\ref{ADGP}) is given by, 
\begin{align}\label{EZGPcov}
\begin{split}
 \phi(\bm{w}_i,\bm{w}_j)&=\hbox{Cov}(Y(\bm{w}_i),Y(\bm{w}_j))\\
    &=\phi_0(\bm{x}_i,\bm{x}_j|\bm{\theta}_0)+\sum_{h=1}^{q}\phi_h(\bm{x}_i,\bm{x}_j|\bm{\Theta}^{(h)})\\
    &=\sigma_0^2\exp\{-\sum_{k=1}^{p}\theta_k^{(0)}(x_{ik}-x_{jk})^2\}\\
    &\ \ \ \ +\sum_{h=1}^{q}\sum_{l_h=1}^{m_h}\mathds{1}(z_{ih}=z_{jh}\equiv l_h)\sigma_h^2\exp\{-\sum_{k=1}^{p}\theta^{(h)}_{kl_h}(x_{ik}-x_{jk})^2\}.
    \end{split}
\end{align}
Given the covariance function, there are  $1+p+q+p\sum_{h=1}^{q}m_h$ parameters $\mu$, $\sigma_0^2$, $\theta_k^{(0)}$, $\sigma_h^2$, and $\theta_{kl_h}^{(h)}$ for  $h=1,\dots,q, k=1,\dots,p$ and $l_h=1,\dots,m_h$. They can be estimated using the maximum likelihood estimation (MLE) method. After dropping the constant terms of the log-likelihood function, under the GP model in (\ref{ADGP}), the MLE method aims to minimize,
\begin{align*}
   \log|\bm{\Phi}|+(\bm{y}-\mu \mathds{1})^{T}\bm{\Phi}^{-1}(\bm{y}-\mu \mathds{1}),
\end{align*} 
with $\mathds{1}$ being a column of $n$ ones, $\bm{\sigma}^2=(\sigma_0^2,\dots,\sigma^2_q)^T$ and $\bm{\Theta}=(\bm{\theta}^{(0)},\bm{\Theta}^{(1)},\dots,\bm{\Theta}^{(q)})$ where $\bm{\theta}^{(0)}=(\theta_k^{(0)})_{ p \times 1}$, $\bm{\Theta}^{(h)}=(\theta_{kl_h}^{(h)})_{p\times m_h}$, $\bm{y}=(y_1,\dots,y_n)^T$ and $\bm{\Phi}=(\phi(\bm{w}_i,\bm{w}_j))_{n\times n}$ with the covariance given in (\ref{EZGPcov}). For a given $\bm{\sigma}^2$ and $\bm{\Theta}$, the maximum likelihood estimate of $\mu$ is, 
\begin{align*}
    \hat{\mu}=(\mathds{1}^T\bm{\Phi}^{-1}\mathds{1})^{-1}\mathds{1}^T\bm{\Phi}^{-1}\bm{y},
\end{align*}
and $\bm{\sigma}^2$ and $\bm{\Theta}$ can be obtained from,
\begin{align*}
    \{\bm{\sigma}^2,\bm{\Theta}\}=\operatorname*{argmin}_{\bm{\sigma}^2,\bm{\Theta}}\{\log|\bm{\Phi}|+\bm{y}^{T}\bm{\Phi}^{-1}\bm{y}-(\mathds{1}^{T}\bm{\Phi}^{-1}\mathds{1})^{-1}(\mathds{1}\bm{\Phi}^{-1}\bm{y})^2\},
\end{align*}
using a minimization algorithm in R or MATLAB. Denote $Y_{*}=Y(\bm{w}^*)$ as the prediction of $Y$ at a new input $\bm{w}^*$. Given $\hat{\mu}$, $\bm{\sigma}^2$ and $\bm{\Theta}$, the predictive mean and the predictive variance at a new location $\bm{w}^{*}$ are given by, 
\begin{align}\label{ARSDC_dist}
\begin{split}
 E(Y_{*}|\bm{y})&=\mu_{*}=\hat{\mu}+\bm{r}_0^{T}\bm{\Phi}^{-1}(\bm{y}-\hat{\mu}\mathds{1}),\\
  \mbox{Var}(Y_{*}|\bm{y})&=\sigma^{{2}}_{*}=\sum_{i=0}^{q}\sigma^2_{i}-\bm{r}_0^T\bm{\Phi}^{-1}\bm{r}_0+\frac{(1-\mathds{1}^T\bm{\Phi}^{-1}\bm{r}_0)^2}{\mathds{1}^T\bm{\Phi}^{-1}\mathds{1}},
    \end{split}
\end{align}
where $\bm{r}_0$ is the covariance vector of $\bm{\Phi}(\bm{w}^{*},\bm{w}_i)_{n\times 1}$ for $i=1,\dots,n$.

\subsection{Acquisition Functions}\label{Section2.2}
This subsection reviews the criteria for  contour estimation  using adaptive designs for 
computer experiments with only quantitative inputs. Adaptive designs start with an initial design with $n_0$ runs. Then based on a fitted model we search for the next input by maximizing a particular criterion $C(\bm{x})$ that will be discussed later. The stopping criterion can be the budget sample size or cost, or when there is little improvement in the prediction accuracy. 

The problem of contour estimation considers finding the inputs that yield a pre-specified contour level of interest.   Suppose $a$ is the contour level of interest and the contour is defined to be $S(a)=\{\bm{x}\in \chi:y(\bm{x})=a\}$ where we let $\chi$ denote the design space throughout. For computer experiments with quantitative inputs,  \cite{ranjan2008sequential} introduced the EI criterion with the improvement function defined as, 
\begin{align}\label{EIran}
    I(\bm{x})=\epsilon^2-\min\{(y(\bm{x})-a)^2,\epsilon^2\},
\end{align}
where $\epsilon=\alpha \hat{\sigma}(\bm{x})$ for $\alpha>0$, and $y(\bm{x})\sim N(\hat{\mu}(\bm{x}),\hat{\sigma}^2(\bm{x}))$ with the predictive mean $\hat{\mu}(\bm{x})$ and the predictive standard deviation $\hat{\sigma}(\bm{x})$. Let ${u}_1=(a-\hat{\mu}(\bm{x})-\epsilon)/\hat{\sigma}(\bm{x})$ and ${u}_2=(a-\hat{\mu}(\bm{x})+\epsilon)/\hat{\sigma}(\bm{x})$. Then the expected value of the improvement function in (\ref{EIran}) is,
\begin{align}\label{ranjan}
\begin{split}
    E[I(\bm{x})]&=[\epsilon^2-(\hat{\mu}(\bm{x})-a)^2-\hat{\sigma}^2(\bm{x})](\Phi({u}_2)-\Phi({u}_1))+\hat{\sigma}^2(\bm{x})(u_2\phi({u}_2)-u_1\phi({u}_1))\\&+2(\hat{\mu}(\bm{x})-a)\hat{\sigma}(\bm{x})(\phi({u}_2)-\phi({u}_1)),
\end{split}
\end{align}
where  $\Phi$ and $\phi$ are the cumulative distribution function (CDF) and the probability density function of the standard normal distribution, respectively. The EI criterion chooses the next input, 
\begin{align*}
    \bm{x}_{n+1}=\operatorname*{argmax}_{\bm{x}\in \chi}{E[I(\bm{x})]}.
\end{align*}
  \cite{cole2023entropy} studied an entropy-based adaptive design for contour estimation. In their study, the problem depends on the definition of the failure region  which is characterized as an output $Y$ exceeding a particular contour level $a$. That is, for a limit state function $g(Y)=Y-a$, such that, 

\begin{align}\label{ECLdef}
    \mathbb{G}=\{\bm{x}\in \chi: g(Y(\bm{x}))>0\}~~\mbox{implying contour}~~ \mathbb{C}=\{\bm{x}\in\chi:g(Y(\bm{x}))=0\},
\end{align} 
where $\mathbb{G}$ is the failure region and $\mathbb{C}$ is the contour of interest. They introduced the entropy-based contour locator (ECL) criterion, 
\begin{align}\label{ELC}
\begin{split}
\medmath{\mbox{ECL}(\bm{x}|g)}=&\medmath{-P(g(Y(\bm{x}))>0)\log P(g(Y(\bm{x}))>0)-P(g(Y(\bm{x}))\leq 0)\log P(g(Y(\bm{x}))\leq 0)}\\
    &\medmath{=-(1-\Phi(\frac{\hat{\mu}(\bm{x})-a}{\hat{\sigma}(\bm{x})}))\log(1-\Phi(\frac{\hat{\mu}(\bm{x})-a}{\hat{\sigma}(\bm{x})}))-\Phi(\frac{\hat{\mu}(\bm{x})-a}{\hat{\sigma}(\bm{x})})\log(\Phi(\frac{\hat{\mu}(\bm{x})-a}{\hat{\sigma}(\bm{x})}))},
\end{split}
\end{align}
where $\Phi$ is the CDF of the standard normal distribution, for any limit state function $g(Y)=b(Y-a)$ where $b\in\{-1,1\}$.  It shall be noted because of symmetry in (\ref{ELC}), the criterion works identically for $g(Y)=b(Y-a)$ where $b\in\{-1,1\}$, meaning that the criterion is unchanged for $b=-1$ or $b=1$.  
The ECL criterion chooses the next input as,
\begin{align*}
    \bm{x}_{n+1}=\operatorname*{argmax}_{\bm{x}\in \chi}\mbox{ECL}(\bm{x}|g).
\end{align*}
Note that both the EI criterion in (\ref{ranjan}) and the ECL criterion in (\ref{ELC}), only depend on the predictive mean and the predictive variance. 

For computer experiments with quantitative and qualitative inputs,  \cite{cai2024adaptive} proposed a sequential design approach for the problem of global optimization. They 
proposed finding the follow-up inputs using the criterion of the lower confidence bound with the adaptive search region for the minimization problem. They named their approach as {\em adaptive-region sequential design} (ARSD). The ARSD  approach chooses the next candidate point to be,
\begin{align*}
    \bm{w}_{n+1}=\operatorname*{argmin}_{\bm{w}\in \mathbb{A}_n}\{\hat{\mu}(\bm{w})-\rho\hat{\sigma}(\bm{w})\},
\end{align*}
where $\mathbb{A}_n \subset \mathbb{A}$ is the adaptive design region as,
\begin{align*}
    \mathbb{A}_n=\{\bm{w}\in \mathbb{A}:\hat{\mu}(\bm{w})-\sqrt{\beta_{0|n}}\hat{\sigma}(\bm{w})\leq \min[\hat{\mu}(\bm{w})+\sqrt{\beta_{0|n}}\hat{\sigma}(\bm{w})]\},
\end{align*}
$\rho \geq 0$ is a tuning parameter, $\mathbb{A}=\{(\bm{x},\bm{z})|\bm{x}\in \chi,\bm{z}\in \bm{Z}\}$ is the whole design space and $\bm{Z}$ is all level combinations of the qualitative inputs, $\hat{\mu}(\bm{w})$ and $\hat{\sigma}(\bm{w})$ are the predicted mean and the predicted standard deviation, $\beta_{0|n}=2\log(\frac{\pi^2n^2M}{6\alpha})$, $\alpha\in(0,1)$ and $M=\prod_{j=1}^{q}m_j$, and $m_j$ is the number of level of the $j$th qualitative input variable.

\section{A  Region-based Cooperative Contour Estimation}
\label{sec:propos}

In this section, we propose an adaptive design approach for the contour estimation problem
when computer experiments involve both quantitative and qualitative input variables.
In principle, any modeling approach that provides the predictive mean and variance can be used in the proposed procedure. For illustrative purposes, we employ the EzGP model by \cite{xiao2021ezgp}. The goal here is to estimate the contour $S(a)=\{\bm{w}\!:y(\bm{w})=a\}$ where $a$ is a prespecified value of the response of interest. That is, the contour estimation is to find the inputs such that the corresponding responses are equal to the contour level $a$.

 Through our empirical investigation, we have observed that the value of the contour level, $a$, plays a critical role in contour estimation because the relative performance of different criteria will change as the value of $a$ changes. In other words, the numerical studies show that at different contour levels, existing criteria provide different comparative results, meaning that one given criterion is not the best choice at all contour levels. Therefore, we propose a method that combines different criteria and performs well in different scenarios. We do this by dividing the design space into two disjoint regions, and apply appropriate criteria in each region. We call the proposed criterion for choosing the next point {\em region-based cooperative
criterion}. In particular, the proposed criterion combines the ideas of the ARSD approach in \cite{cai2024adaptive} and the ECL criterion in \cite{cole2023entropy}.

\subsection{A  Region-based Cooperative
Criterion}

Let the lower bound (LB) and the upper bound (UB) be given by,
\begin{align}\label{LBUB}
\begin{split}
    &\mu_{h}^{L}(\bm{w}_0)=|\hat{\mu}_{0|n}(\bm{w}_0)-a| - \sqrt{\beta_{0|n}}\hat{\sigma}_{0|n}(\bm{w}_0)\\&\mu_{h}^{U}(\bm{w}_0)=|\hat{\mu}_{0|n}(\bm{w}_0)-a| + \sqrt{\beta_{0|n}}\hat{\sigma}_{0|n}(\bm{w}_0),
   \end{split}
    \end{align}
where $\hat{\mu}_{0|n}(\bm{w}_0)$ is the predictive mean, $\hat{\sigma}_{0|n}(\bm{w}_0)$ is the predictive standard deviation defined in (\ref{ARSDC_dist}),  $\beta_{0|n}=2\log(\frac{\pi^2n^2 M}{6\alpha})$, $\alpha\in(0,1)$, $M=\prod_{j=1}^{q}m_j$ and $m_j$ is the number of levels of the $j$th qualitative input variable. Note that, the choice of $\beta_{0|n}$ is from \cite{cai2024adaptive}. Let $\mathbb{A}$ be the whole design space, we now divide $\mathbb{A}$ into two disjoint groups, 
\begin{align}
    &\mathbb{A}_{1}=\{\bm{w}_0\in \mathbb{A}:|\hat{\mu}_{0|n}(\bm{w}_0)-a|-\sqrt{\beta_{0|n}}\hat{\sigma}_{0|n}(\bm{w}_0)>0\}.\label{ACCEA}\\
    &\mathbb{A}_2=\{\bm{w}_0\in \mathbb{A}:|\hat{\mu}_{0|n}(\bm{w}_0)-a|-\sqrt{\beta_{0|n}}\hat{\sigma}_{0|n}(\bm{w}_0) \leq 0\}.\label{A_3}
        \end{align}
Clearly, we have $\mathbb{A}= \cup_{i=1}^2 \mathbb{A}_i$. For $\mathbb{A}_1$ in (\ref{ACCEA}), consider the candidate point given by, 
\begin{align}\label{criterion_min}
        \bm{w}_{n+1}=\operatorname*{argmax}_{\bm{w}_0\in\mathbb{A}_{1,\min}}\hat{\sigma}_{0|n}(\bm{w}_0),
\end{align}
where,
\begin{align}\label{A_min}
\medmath{\mathbb{A}_{1,\min}=\{\bm{w}_0\in \mathbb{A}_{1}:|\hat{\mu}_{0|n}(\bm{w}_0)-a|-\sqrt{\beta_{0|n}}\hat{\sigma}_{0|n}(\bm{w}_0) \leq \min_{\bm{w}_0\in\mathbb{A}}[|\hat{\mu}_{0|n}(\bm{w}_0)-a|+\sqrt{\beta_{0|n}}\hat{\sigma}_{0|n}(\bm{w}_0)] \}.}
 \end{align}
For $\mathbb{A}_2$ in (\ref{A_3}), we let the candidate point be,
\begin{align}\label{criterion_A_3}
    \bm{w}_{n+1}=\operatorname*{argmax}_{\bm{w}_0\in \mathbb{A}_{2}} \mbox{ECL}[\bm{w}_0|g],
\end{align}
with the ECL criterion in (\ref{ELC}).


Among the two candidate points selected from groups $\mathbb{A}_1$ and $\mathbb{A}_2$, we select the next input with a larger value of $\frac{\hat{\sigma}_{0|n}(\bm{w}_0)}{\max(\delta,|\hat{\mu}_{0|n}(\bm{w}_0)-a|)}$, where $\delta$ is a small positive value. The choice of $\delta$ determines the degree of exploration and exploitation. A very small value of $\delta$ favors exploitation and chooses inputs mostly, if not exclusively, from $\mathbb{A}_2$. A very large value of $\delta$ would be equivalent to using the predictive variance only and results only exploration, leading to choosing excessive inputs from $\mathbb{A}_1$. To balance exploration and exploitation, the value of $\delta$ should not be too large or too small.   In the numerical examples and the real application, we will give the recommended choice of $\delta$ which depends on the range of responses.  In general, we aim to select the point whose predictive mean is closer to the contour level $a$ with the larger predictive variance. However, such a choice will empirically result in the selection from only one group after some steps in the sequential procedure. To promote the exploration of the design space, we use ${\max(\delta,|\hat{\mu}_{0|n}(\bm{w}_0)-a|)}$ in the denominator so that we choose a point with a larger value of  $\hat{\sigma}_{0|n}(\bm{w}_0)$ when the predictive means of both candidate points are close to the contour level. 
 Algorithm \ref{alg:cap2} describes the proposed adaptive design approach for finding follow-up points for contour estimation.  In a nutshell, we apply the idea of the adaptive search region in the ARSD approach in the design space $\mathbb{A}_{1}$, and the ECL criterion in the design space $\mathbb{A}_{2}$. The difference between our approach and the ARSD approach is we search for the candidate point that has the largest predictive uncertainty while the ARSD approach aims to minimize the lower confidence bound in solving the minimization problem. One advantage of only applying the ECL criterion in $\mathbb{A}_{2}$ is that the input that maximizes the criterion is unique, as seen in our empirical studies.  In contrast, if applying the ECL criterion in the whole design space with mixed inputs, there are several candidate inputs maximizing the criterion. This advantage also results in the fact that 
 using the limit state function $g(y) = y-a$ or $g(y) = -(y-a)$ in our proposed procedure gives the same input selection.

 It shall be noted that $\mathbb{A}_{1}$ contains the design points that the contour level $a$ is outside the confidence band $(\hat{\mu}_{0|n}(\bm{w}_0)- \sqrt{\beta_{0|n}}\hat{\sigma}_{0|n}(\bm{w}_0), \hat{\mu}_{0|n}(\bm{w}_0)+\sqrt{\beta_{0|n}}\hat{\sigma}_{0|n}(\bm{w}_0) )$ while  $\mathbb{A}_{2}$ consists of the design points that the contour level $a$ is inside the confidence band. If using ECL alone,  $\mathbb{A}_{2}$ is where most likely the next follow-up point will be from. Borrowing the success of ECL, we apply it in the region  $\mathbb{A}_{2}$. The search in  $\mathbb{A}_{1}$ represents the effort of exploration that is needed for any adaptive design. As shown in the numerical example, in the beginning of the adaptive design process, the follow-up point is from $\mathbb{A}_{2}$ only, and as more points are added, the exploration is taking effect, and more follow-up inputs are from $\mathbb{A}_{1}$. 
We call the proposed method  {\em region-based cooperative
criterion} (RCC) in the rest of the paper. 
\begin{algorithm}
\caption{Region-based Cooperative Contour Estimation}
\label{alg:cap2}
\begin{algorithmic}[1]  
\State Run a small initial design $\bm{W}_{n_0}=(\bm{w}^T_1,\dots, \bm{w}^T_{n_0})$.
\State Obtain $\bm{D}_{n_0}=(\bm{W}_{n_0},\bm{Y}_{n_0})$, where $\bm{Y}_{n_0}=(f(\bm{w}_{1}),\dots,f(\bm{w}_{n_0}))$.
\State Set $n = n_0$.
\While{$n < N$}
    \State Fit the EzGP model using $\bm{D}_n$.
    \State Divide the search space into two groups: $\mathbb{A}_1$ and $\mathbb{A}_2$.
    \State For $\mathbb{A}_1$ (defined in (\ref{ACCEA})), find the input minimizing the criterion in (\ref{criterion_min}).
    \State For $\mathbb{A}_2$ (defined in (\ref{A_3})), find the input maximizing the criterion in (\ref{criterion_A_3}).
    \State Choose the input from the two chosen points that $\frac{\hat{\sigma}_{0|n}(\bm{w}_0)}{\max(\delta,|\hat{\mu}_{0|n}(\bm{w}_0)-a|)}$ is larger.
    \State Run the experiment at $\bm{w}_{n+1}$ and obtain $y_{n+1}=f(\bm{w}_{n+1})$.
    \State Update $\bm{D}_{n+1} = \bm{D}_n \cup (\bm{w}_{n+1}, y_{n+1})$, set $n \gets n+1$.
\EndWhile
\State \Return Extract the estimated contour $\hat{S} = \{\bm{w}_0 : \hat{y}(\bm{w}_0) = a\}$.
\end{algorithmic}
\end{algorithm}

\subsection{Theoretical Justification}\label{section3.2}

This subsection provides the theoretical justifications of the adaptive search region.
These justifications are in a similar spirit to those provided by \cite{cai2024adaptive}. The differences between ours and theirs are two-fold: (1) theirs are for the global optimization in computer experiments with quantitative and qualitative factors while ours are for the contour estimation; (2) their adaptive search region does not involve with grouping while ours does. To accommodate the first difference, their justifications give the theoretical guarantee for the minimum of $Y$, while ours are for the minimum of $|Y(\bm{w}_0)-a|$.  For notational convenience, we use $\mu_{0|n}$ and $\sigma_{0|n}$ instead of the estimation for the statements and proofs. Let $Y(\bm{w}_0)$ be the response function, and assume $Y$ is a sample of a GP. The covariance function considered can be any covariance function that can handle mixed inputs.

\begin{lemma}\label{lemma1}
For a given input $\bm{x}_0\in \chi$ for quantitative factors that gives a design point $\bm{w}_0=(\bm{x}_0^T,\bm{z}_0^T)^T\in \mathbb{A}$, consider $h(\bm{w}_0)=|Y(\bm{w}_0)-a|$ and $\mu_h(\bm{w}_0)=|\mu_{0|n}(\bm{w}_0)-a|$. For all $\alpha\in(0,1)$, we have, 
  \begin{displaymath}
   P(|h(\bm{w}_0)-\mu_h(\bm{w}_0)| \leq\sqrt{\beta_{0|n}}\sigma_{0|n}(\bm{w}_0),\forall \bm{z}_0\in \bm{Z}, \forall \ n\geq 1)>1-\alpha,
  \end{displaymath}
where $\bm{Z}$ represents all level combinations of the qualitative inputs, $\mu_{0|n}(\bm{w}_0)$ and $\sigma_{0|n}(\bm{w}_0)$ are the predictive mean and the predictive standard deviation, respectively, and $\beta_{0|n}=2\log(\frac{\pi^2n^2M}{6\alpha})$ with $M=|\bm{Z}|=\prod_{j=1}^{q}m_j$ being the size of $\bm{Z}$.
\end{lemma}
The proof is stated in the Appendix. Note that Lemma \ref{lemma1} applies for any value of $a$. Before stating Lemma \ref{lemma2}, we first need to define some notations. Let $\mu_h(\bm{w}_0)=|\mu_{0|n}(\bm{w}_0)-a|$,  
\begin{align}\label{lemma2def}
    \begin{split}
    &h_{\min}=\min_{\bm{w}_0\in \mathbb{A}}h(\bm{w}_0)=\min_{\bm{w}_0\in\mathbb{A}}|Y(\bm{w}_0)-a|;\\&\tilde{\mu}_{\min,n}=\min_{\bm{w}_0\in \mathbb{A}}\mu_{h}(\bm{w}_0)=\min_{\bm{w}_0\in\mathbb{A}}|\mu_{0|n}(\bm{w}_0)-a|;\\
        &\tilde{\mu}^{L}_{\min,n}=\min_{\bm{w}_0\in\mathbb{A}}\mu_{h}^{L}(\bm{w}_0);~
        \tilde{\mu}^{U}_{\min,n}=\min_{\bm{w}_0\in\mathbb{A}}\mu_{h}^{U}(\bm{w}_0)
    \end{split}
\end{align}
where $\mu_{h}^{L}(\bm{w}_0)$ and $\mu_{h}^{U}(\bm{w}_0)$ are defined in (\ref{LBUB}). Lemma \ref{lemma2} below states that with a probability greater than $1-\alpha$$,~ h_{\min}$ is in the interval with the lower bound $\tilde{\mu}^{L}_{\min,n}$ and the upper bound $\tilde{\mu}^{U}_{\min,n}$ defined in (\ref{lemma2def}) for any value of $a$.
\begin{lemma}\label{lemma2}
Let $h_{\min}$, $\tilde{\mu}^{L}_{\min,n}$ and $\tilde{\mu}^{U}_{\min,n}$ defined in (\ref{lemma2def}), let $a$ be the contour level, then for all $\alpha\in(0,1)$, 
  \begin{displaymath}
P(h_{\min}\in[\tilde{\mu}^{L}_{\min,n},\tilde{\mu}^{U}_{\min,n}], \forall \ n\geq 1)> 1-\alpha.
  \end{displaymath}
\end{lemma}

The proof is stated in the Appendix. Using Lemmas \ref{lemma1} and \ref{lemma2}, we have Theorem \ref{Theorem1}.
\begin{theorem}\label{Theorem1} Let  $h_{\min}$, $\tilde{\mu}_{min,n}$, $\tilde{\mu}^{L}_{\min,n}$,  $\tilde{\mu}^{U}_{\min,n}$, $\mu^{L}_{h}$ and $\mu^{U}_{h}$ defined in (\ref{LBUB}) and (\ref{lemma2def}). Then let $a$ be the contour level, then for all $\alpha\in(0,1)$, 
  \begin{displaymath}
    P(|\tilde{\mu}_{\min,n}-h_{\min}|\leq  \sqrt{{\beta}_{0|n}} \sup_{\bm{w}_0\in\mathbb{A}_{c}}\sigma_{0|n}(\bm{w}_0),\forall \ n\geq 1)> 1-\alpha,
\end{displaymath}
 where $\mathbb{A}_{c}=\mathbb{A}_{1,\min}\cup\mathbb{A}_{2}$ with  $\mathbb{A}_{2}$  and $\mathbb{A}_{1,\min}$ given in (\ref{A_3}) and (\ref{A_min}).
\end{theorem} 
Theorem \ref{Theorem1} provides a bound for the difference of $\tilde{\mu}_{\min,n}$ and $h_{\min}$ that depends on $\mathbb{A}_{c}=\mathbb{A}_{1,\min}\cup\mathbb{A}_{2}$, and it states that with a probability higher than $1-\alpha$, $\tilde{\mu}_{\min,n}$ will be in the neighborhood of $h_{\min}$.

\section{Numerical Experiments}\label{sec:simul}

In this section, we use numerical examples to illustrate the effectiveness of the proposed approach. We compare its performance with the major competitors including EI, ECL, the lower confidence bound (LCB) criterion, and one-shot designs. The LCB criterion chooses the next input as,
\begin{align*}
        \bm{w}_{n+1}=\operatorname*{argmin}_{\bm{w}_0\in\mathbb{A}}\{|\hat{\mu}_{0|n}(\bm{w}_0)-a|-\rho\hat{\sigma}_{0|n}(\bm{w}_0)\},
\end{align*}
where $\rho$ is a tuning parameter and $\mathbb{A}$ is the whole design space. Note that the LCB criterion was proposed for the problem of global optimization, but we apply it to the contour estimation problem by using $|\hat{\mu}_{0|n}(\bm{w}_0)-a|$ rather than $\hat{\mu}_{0|n}$. Similarly,  the ARSD approach for optimization can also be modified for contour estimation. That is, we can choose the next input to be, 
\begin{align*}
        \bm{w}_{n+1}=\operatorname*{argmin}_{\bm{w}_0\in\mathbb{A}^{*}}\{|\hat{\mu}_{0|n}(\bm{w}_0)-a|-\rho\hat{\sigma}_{0|n}(\bm{w}_0)\},
\end{align*}
where $\rho$ is a tuning parameter, and 
\begin{align*}
 \mathbb{A}^{*}=\{\bm{w}_0\in \mathbb{A}:|\hat{\mu}_{0|n}(\bm{w}_0)-a|-\sqrt{\beta_{0|n}}\hat{\sigma}_{0|n}(\bm{w}_0) \leq \min_{\bm{w}_0\in\mathbb{A}}[|\hat{\mu}_{0|n}(\bm{w}_0)-a|+\sqrt{\beta_{0|n}}\hat{\sigma}_{0|n}(\bm{w}_0)] \}.
 \end{align*}
We denote this approach as ARSD. We also consider a criterion that replaces the ECL criterion in (\ref{criterion_A_3}) with the EI criterion in (\ref{ranjan}), and denote the criterion as RCC-EI. 
In all adaptive design approaches, we use $n_0$-run initial designs, and find the next input by considering a large number of candidate points in the design space.  Here we let the candidate points be all level combinations of the qualitative inputs and  100 random Latin hypercube designs (LHD) \cite{r-core} for each level combination in the design space. One-shot designs  consist of random LHDs  for the quantitative variables and a (nearly) balanced sample from all level combinations for the qualitative factors. The random LHDs are generated using the $randomLHS$ in the R package $lhs$ \cite{r-core}.  To determine the tuning parameter $\rho$, we compare different values of $\rho=0.5,1,2$. We choose $\rho=2$ based on simulation results as this value returns the best results. 

To evaluate the efficiency of these criteria in the simulation studies, we consider 
\begin{align}\label{mc0}
M_{C_{0}}=\frac{1}{|C_0|}\sum_{\bm{w}_0\in C_0 }|Y(\bm{w}_0)-\hat{Y}(\bm{w}_0)|,
\end{align}
 where $C_0=\{\bm{w}_0:Y(\bm{w}_0)=a\}$ is extracted from a large set of candidate points satisfying $a\pm \epsilon$ where $\epsilon$ is a small positive value and it has been chosen based on the response function's complexity to ensure there are enough points to calculate the measurement. The measurement $M_{C_{0}}$ measures the average discrepancy between the true function values and the predicted means and thus, the smaller value of $M_{C_0}$ indicates that the fitted surface around the true contour is a more accurate approximation to the true surface and thus is preferred. The size of the large set of candidate points (random LHDs) depends on the number of all level combinations of the qualitative factors. In all the examples below, we set the size to be 200 random LHDs for each level combination to calculate $M_{C_0}$. In addition to $ M_{C_0}$ in (\ref{mc0}), we have tried different measurements such as sensitivity and specificity (\cite{cole2023entropy}). Sensitivity is the proportion of the input space that is correctly identified as being inside of the failure region that has been defined in (\ref{ECLdef}), and specificity is the proportion of the input space that is correctly identified as being outside the failure region. However, for contour estimation with mixed inputs, it is not the sensible measurement as it does not represent the accuracy of contour estimation well. This is because in our case the failure region defined in (\ref{ECLdef}) often involves different level combinations of qualitative factors, and the model fitting at those level combinations may not be accurate but the design space associated with those level combinations are identified as the correct failure region.  Thus, the sensitivity and specificity criteria are not included for contour estimation when we have quantitative and qualitative input variables. 
 
\subsection*{Example 1}\label{ex1}
\begin{figure}[htbp]
\centering
\includegraphics[scale=0.5]{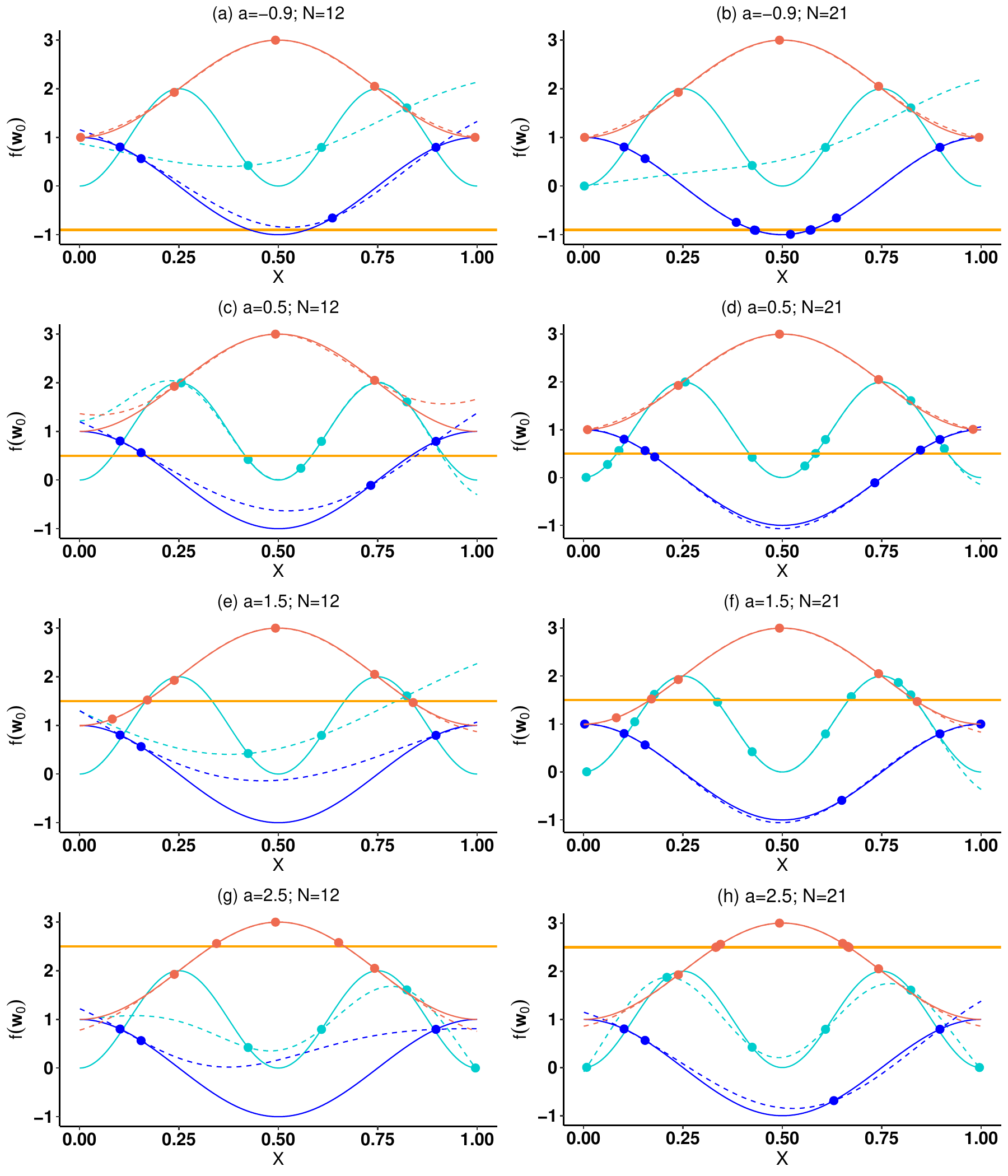}
\caption{\footnotesize The adaptive designs using the RCC criterion for  \nameref{ex1} with $n_0=9$ and for the contour level $a=\{-0.9,0.5,1.5,2.5\}$, for the total sample sizes of $N=\{12,21\}$. The solid lines represent the underlying true response and the dotted lines represent the predicted mean. The orange line shows the contour level.}\label{stages_ex1}
\end{figure}

Consider a computer experiment with $p=1$ quantitative input $x$ and $q=1$ qualitative input $z$ and the computer model is represented by, 
\begin{align*}
  f=\begin{cases}
      2-\cos(2\pi {x}), & \text{if}\ {z}=1\\  1-\cos(4\pi {x}), & \text{if}\ {z}=2\\ \cos(2\pi {x}), &  \text{if}\ {z}=3.
    \end{cases}
\end{align*}
We consider the contour levels $a=\{-0.9,0.5,1.5,2.5\}$. Here $-1$ is the minimum of the response surface and $3$ is the maximum of the response surface. We let  $n_0=9$, $N=\{12,15,18,21\}$, $\epsilon=0.05$ for the measurement $M_{C_0}$ in (\ref{mc0}) and $\delta =0.05$. Figure \ref{stages_ex1} displays one adaptive design using the proposed RCC, for $a=\{-0.9,0.5,1.5,2.5\}$ and reveals that the next input is chosen close to $a$ using RCC. Table \ref{Table_ex1} reports the average of the measurements $M_{C_{0}}$ using the one-shot designs and the adaptive designs over 50 simulations for different contour levels. The values in the parentheses are the relative efficiency of the criteria over one-shot designs. The table reveals that indeed the adaptive designs generally provide more accurate contour estimation than one-shot designs. All the adaptive design approaches yield the smallest value of the average of the measurements $M_{C_{0}}$ 
at certain scenarios, and none of the methods provide the most accurate estimations at all contour levels with all the sample sizes. However, out of all adaptive designs with different contour levels and different sample sizes, most of the time the relative efficiency of the proposed method is higher than other methods, or close to the other criteria we compare with. Especially when the run size gets larger, there is marked improvement of the relative efficiency of the proposed approach. Table \ref{Ex1-time} shows the average computation time to add 12 points adaptively using each adaptive design. It can be observed that the computation time in seconds of the ARSD and the RCC methods are much less than  that of ECL, EI, and LCB. 

\begin{table}[H]
\footnotesize
\caption{\footnotesize The average of the measurements $M_{C_{0}}$ in Example 1 for the adaptive designs ARSD, RCC, RCC-EI, ECL, EI, LCB, and one-shot designs for $n_0=9$ and $N=\{12,15,18,21\}$ over 50 simulations for the contour level $a=\{-0.9,0.5,1.5,2.5\}$. The values in parentheses are the relative efficiency of the criteria over one-shot designs.}
\label{Table_ex1}
\vspace{10pt}
\resizebox{15cm}{!}{
\setlength\tabcolsep{2pt}
  \begin{tabular}{r rrrrrrrrr}
    \hline 
  $N$~~~\textbf{one-shot} & \textbf{ARSD} &\textbf{{RCC}}& \textbf{{RCC-EI}}&\textbf{ECL} & \textbf{EI}& \textbf{LCB}&\\
   \hline
 \hline& &&{a=-0.9}&&&\\
   \hline
   12~~~~0.0865 &0.0508(1.70)& 0.1524(0.57)& 0.2141(0.40)& 0.1491(0.58)& 0.2141(0.40)&\textbf{0.0506(1.71)}\\ 
  15~~~~0.0197 & 0.0023(8.57)&\textbf{0.0006(32.8)}& 0.0627(0.31)& 0.0007(28.1)& 0.0411(0.48)& 0.0023(8.57)\\ 
18~~~~0.0091 & \textbf{0.0001(91.0)}&\textbf{0.0001(91.0)}& 0.0005(18.2)& \textbf{0.0001(91.0)}& 0.0005(18.2)&\textbf{0.0001(91.0)}\\ 
21~~~~0.0020 &\textbf{0.0001(20.0)}&\textbf{0.0001(20.0)}&\textbf{0.0001(20.0)}& \textbf{0.0001(20.0)}& \textbf{0.0001(20.0)}&\textbf{0.0001(20.0)}\\     
    \hline& &&{a=0.5}&&&\\
   \hline
12~~~~0.4533 & 0.4394(1.03)& 0.4004(1.13)& 0.4047(1.12)& \textbf{0.3895(1.16)}&0.4047(1.12)& 0.4367(1.04)\\ 
 15~~~~0.3124 & 0.1845(1.69)& 0.2065(1.51)& 0.1978(1.58)& 0.2370(1.32)& 0.1978(1.58)&\textbf{0.1812(1.72)}\\ 
  18~~~~0.1456 & 0.0465(3.13)&\textbf{0.0319(4.56)}& 0.0381(3.82)& 0.0653(2.23)& 0.0343(4.24)& 0.0469(3.10)\\ 
  21~~~~0.0562 & 0.0012(46.8)&\textbf{0.0007(80.3)}& 0.0029(19.3)& 0.0148(3.80)& 0.0030(18.73)& 0.0012(46.8)\\
\hline& &&{a=1.5}&&&\\
   \hline
   12~~~~0.3608 & 0.4634(0.78)& \textbf{0.4120(0.88)}& 0.4766(0.76)& 0.4557(0.79)& 0.4766(0.76)& 0.4688(0.77)\\ 
 15~~~~0.2219 & 0.2635(0.84)&\textbf{0.1879(1.18)}& 0.2169(1.02)& 0.2059(1.08)& 0.2169(1.02)& 0.2745(0.81)\\ 
  18~~~~0.1137 & 0.0390(2.92)&0.0396(2.87)& 0.0261(4.36)& 0.0423(2.69)&\textbf{0.0261(4.36)}& 0.0398(2.86)\\ 
  21~~~~0.0392 & 0.0011(35.6)& \textbf{0.0005(78.4)}& 0.0018(21.8)& 0.0009(43.6)& 0.0018(21.8)& 0.0011(35.6)\\     
 \hline& &&{a=2.5}&&&\\
   \hline
 12~~~~0.1343 & 0.1004(1.34)& 0.0510(2.63)& 0.1051(1.28)& \textbf{0.0234(5.74)}& 0.1051(1.28)& 0.1004(1.34) \\ 
 15~~~~0.0248 & 0.0038(6.53)& 0.0006(41.33)& 0.0290(0.86)& \textbf{0.0005(49.6)}& 0.0291(0.85)& 0.0038(6.53) \\ 
  18~~~~0.0074 & \textbf{0.0001(74.0)}& \textbf{0.0001(74.0)}& 0.0004(18.5)& \textbf{0.0001(74.0)}& 0.0006(12.3)& \textbf{0.0001(74.0)}\\ 
  21~~~~0.0034 & \textbf{0.0001(34.0)}& \textbf{0.0001(34.0)}& \textbf{0.0001(34.0)}& \textbf{0.0001(34.0)}& \textbf{0.0001(34.0)}& \textbf{0.0001(34.0)}\\
    
    \end{tabular}
    }
  \end{table}

\begin{table}[H]
\footnotesize
   \centering
\caption{\footnotesize The average computation time (in seconds) in \nameref{ex1} for the adaptive designs ARSD, RCC, RCC-EI, ECL, EI, and LCB for $n_0=9$ and $N=21$ over 50 simulations for the contour level $a=\{-0.9,0.5,1.5,2.5\}$.}\label{Ex1-time}
\vspace{10pt}
\resizebox{9cm}{!}{
\setlength\tabcolsep{2pt}
  \begin{tabular}{r rrrrrrrrrrrr}
    \hline 
  a~~~~~&&\textbf{ARSD}& &\textbf{{RCC}}& &\textbf{{RCC-EI}}& &\textbf{ECL} & &\textbf{EI}& &\textbf{LCB}\\[0.05cm]  \hline
   -0.9~~~~~&&0.16& &0.21& &0.24& &1.50 & &1.54& &1.47\\
   0.5~~~~~&&0.22& &0.36& &0.40& &1.46 & &1.49& &1.54\\
   1.5~~~~~&&0.22& &0.34& &0.74& &1.58 & &1.49& &1.71\\
   2.5~~~~~&&0.18& &0.21& &0.26& &1.48 & &1.61& &1.57\\
      \end{tabular}
      }
  \end{table}
  
\subsection*{Example 2}\label{ex2}
We consider a computer model with $p=2$ quantitative inputs $\bm{x}=(x_1,x_2)$ and $q=2$ qualitative input $\bm{z}=(z_1,z_2)$. The computer model is adapted from \cite{xiao2021ezgp} with reduced dimensions and is  represented by, 
\begin{align*}
  i({z}_1)=
    \begin{cases}
      x_1+x_2^2, & \text{if}\ z_1=1 \\
      x_1^2+x_2, & \text{if}\ z_1=2\\
      x_1^2+x_2^2,&  \text{if}\ z_1=3,
    \end{cases}~~
  g({z}_2)=
    \begin{cases}
     \cos( x_1)+\cos(2 x_2), & \text{if}\ z_2=1 \\
      \cos( 2x_1)+\cos(x_2), & \text{if}\ z_2=2\\
     \cos(2x_1)+\cos(2x_2),&  \text{if}\ z_2=3,
    \end{cases}\\
    f=i(z_1)+g(z_2).
\end{align*}
We consider the contour levels of interest $a=\{1.1,1.5,2,2.6\}$. Here 1 is the minimum of the response surface and 2.7 is the maximum of the response surface. We let $n_0=9$, $N=\{27,36,45,54\}$, $\epsilon=0.05$  for the measurement $M_{C_0}$ in (\ref{mc0}),  and $\delta =0.02$. Table \ref{Table_ex2} displays the average of the measurements $M_{C_{0}}$ using the one-shot designs and the adaptive designs over 50 simulations for different contour levels and $N=\{27,36,45,54\}$. The values in the parentheses are the relative efficiency of the criteria over one-shot designs. Again out of all adaptive designs and different contour levels, most of the time the relative efficiency of the proposed method is higher than other methods, or close to the other criteria we compare with. Table \ref{Ex2-time} shows the average computation time to add 45 points adaptively using each adaptive design. It can again be observed that the computation time in seconds of the ARSD and the RCC methods are far less than that of ECL, EI, and LCB.

\begin{table}[!htp]
    \centering
\caption{\footnotesize  The average of the measurements $M_{C_{0}}$ in \nameref{ex2} for the adaptive designs ARSD, RCC, RCC-EI, ECL, EI, LCB and one-shot designs for $n_0=9$ and $N=\{27,36,45,54\}$ over 50 simulations for the contour level $a=\{1.1,1.5,2,2.6\}$. The values in the parentheses are the relative efficiency of the criteria over one-shot designs.}\label{Table_ex2}
\vspace{10pt}
\resizebox{15cm}{!}{
\setlength\tabcolsep{2pt}
  \begin{tabular}{r rrrrrrrrr}
    \hline 
  $N$~~~\textbf{one-shot} & \textbf{ARSD} &\textbf{{RCC}}& \textbf{{RCC-EI}}& \textbf{ECL} & \textbf{EI}& \textbf{LCB}&\\[0.05cm]
   \hline
 \hline& &&{a=1.1}&&&\\
   \hline
   27~~~~0.0966&0.0143(6.76) &0.0105(9.20)&0.0220(4.39)& \textbf{0.0100(9.66)}& 0.0191(5.06)&0.0122(7.92)\\
    36~~~~0.0486&0.0034(14.3) &\textbf{0.0010(48.6)}&0.0028(17.4)& 0.0017(28.6)& 0.0043(11.3)&0.0034(14.3)\\
  45~~~~0.0301&0.0007(43.0) &\textbf{0.0003(100)}&\textbf{0.0003(100)}& 0.0005(60.2)& \textbf{0.0003(100)}&0.0006(50.1)\\
 54~~~~0.0092&0.0003(30.7) &\textbf{0.0001(92.0)}&{0.0002(46.0)}& 0.0002(46.0)& 0.0002(46.0)&0.0003(30.7)\\
    \hline& &&{a=1.5}&&&\\
   \hline
    27~~~~0.0493&0.0225(2.19) & \textbf{0.0143(3.45)}&0.0335(1.39)& 0.0155(3.18)& 0.0372(1.33)&0.0268(1.84)\\
    36~~~~0.0245&0.0075(3.27) & {0.0035(7.00)}&0.0117(2.09)& \textbf{0.0031(7.90)}& 0.0102(2.40)&0.0072(3.40)\\
 45~~~~0.0100&0.00010(10.0) & \textbf{0.0007(14.3)}&0.0017(5.88)& \textbf{0.0007(14.3)}& 0.0020(5.00)&0.0011(9.09)\\
 54~~~~0.0045&0.0003(15.0) & \textbf{0.0002(22.5)}&{0.0003(15.0)}& \textbf{0.0002(22.5)}& 0.0005(9.00)&0.0003(15.0)\\
    \hline& &&{a=2}&&&\\
   \hline
   27~~~~0.0523&0.0543(0.96) & \textbf{0.0464(1.13)}&0.0576(0.91)& {0.0479(1.09)}& 0.0562(0.93)&0.0516(1.01)\\
    36~~~~0.0238&0.02000(1.19) & \textbf{0.0146(1.63)}&0.0211(1.13)& {0.0152(1.57)}& 0.0201(1.18)&0.0181(1.31)\\
  45~~~~0.0099&0.0050(1.98) & \textbf{0.0039(2.54)}&0.0052(1.90)& 0.0040(2.48)& 0.0051(1.94)&0.0050(1.98)\\
 54~~~~0.0040&0.0015(2.67) & \textbf{0.0012(3.33)}&0.0016(2.50)& 0.0014(2.86)& 0.0014(2.86)&\textbf{0.0012(3.33)}\\
 \hline& &&{a=2.6}&&&\\
 \hline
    27~~~~0.0977&0.0452(2.16) & 0.0362(2.70)&0.0506(1.93)& \textbf{0.0279(3.50)}& 0.0423(2.31)&0.0604(1.62)\\
    36~~~~0.0385&0.0104(3.70) & \textbf{0.0011(35.0)}&0.0040(9.62)& {0.0018(21.4)}& 0.0027(14.3)&0.0092(4.18)\\
  45~~~~0.0150&0.0011(13.6) & \textbf{0.0003(50.0)}&{0.0004(37.5)}& \textbf{0.0003(50.0)}& 0.0006(25.0)&0.0007(21.4)\\
 54~~~~0.0061&\textbf{0.0002(30.5)}& \textbf{0.0002(30.5)}&\textbf{0.0002(30.5)}& \textbf{0.0002(30.5)}& 0.0004(15.2)&0.0003(20.3)\\  
    \end{tabular}
    }
  \end{table}

 \begin{table}[!htp]
\footnotesize
   \centering
\caption{\footnotesize The average computation time per second in \nameref{ex2} for the adaptive designs ARSD, RCC, RCC-EI, ECL, EI, and LCB  for $n_0=9$ and $N=54$ over 50 simulations for the contour level $a=\{1.1,1.5,2,2.6\}$.}  \label{Ex2-time}
\vspace{10pt}
\resizebox{9cm}{!}{
\setlength\tabcolsep{2pt}
  \begin{tabular}{r rrrrrrrrrrrr}
    \hline 
  $a$~~~~~&&\textbf{ARSD}& &\textbf{{RCC}}& &\textbf{{RCC-EI}}& &\textbf{ECL} & &\textbf{EI}& &\textbf{LCB}\\[0.05cm]  \hline
   1.1~~~~~&&9.23& &7.70& &6.32& &37.3 & &35.1& &35.5\\
   1.5~~~~~&&8.44& &9.44& &7.67& &36.5 & &35.9& &35.3\\
   2~~~~~&&9.99& &10.4& &11.2& &35.5 & &35.1& &37.8\\
   2.6~~~~~&&8.31& &7.98& &6.90& &35.2 & &37.8& &35.9\\
      \end{tabular}
      }
  \end{table}
  
\subsection*{Example 3}\label{ex3}
We consider a computer model with $p=3$ quantitative inputs $\bm{x}=(x_1,x_2,x_3)$ and $q=3$ qualitative input $\bm{z}=(z_1,z_2,z_3)$ and the computer model \cite{xiao2021ezgp} is represented by, 
\begin{align*}
  i({z}_1)=
    \begin{cases}
      x_1+x_2^2+x_3, & \text{if}\ z_1=1 \\
      x_1^2+x_2+x_3, & \text{if}\ z_1=2\\
      x_3+x_1+x_2^2,&  \text{if}\ z_1=3,
    \end{cases}~~
  g({z}_2)=
    \begin{cases}
     \cos( x_1)+\cos(2 x_2)+\cos(x_3), & \text{if}\ z_2=1 \\
      \cos( x_1)+\cos(2 x_2)+\cos(x_3), & \text{if}\ z_2=2\\
     \cos(2x_1)+\cos(x_2)+\cos(x_3),&  \text{if}\ z_2=3,
    \end{cases}
\end{align*}
\begin{align*}
  h({z}_3)=
    \begin{cases}
       \sin(x_1)+\sin(2 x_2)+\sin(x_3), & \text{if}\ z_3=1 \\
      \sin(x_1)+\sin(2 x_2)+\sin(x_3), & \text{if}\ z_3=2\\
     \sin(2x_1)+\sin(x_2)+\sin(x_3),&  \text{if}\ z_3=3,
    \end{cases}~~
    f=i(z_1)+g(z_2)+h(z_3).
\end{align*}
We consider the contour levels of interest $a=\{4,5,6,6.6\}$. Here 3 is the minimum of the response surface and 6.7 is the maximum of the response surface. We let  $n_0=9$, $N=\{27,36,45,54\}$, $\epsilon=0.1$  for the measurement $M_{C_0}$ in (\ref{mc0}), and $\delta =0.1$. Table \ref{Table_ex3} displays the average of the measurements $M_{C_{0}}$ using the one-shot designs and the adaptive designs over 50 simulations for different contour levels and $N=\{27,36,45,54\}$. The values in the parentheses are the relative efficiency of the adaptive designs over one-shot designs. Again out of all adaptive designs and different contour levels, most of the time the relative efficiency of the proposed method is higher than other methods, or close to the other criteria we compare with. Note that using the measurement $M_{C_0}$, the ECL criterion does not always provide comparative results. Table \ref{Ex3-time} shows the average computation time to add 45 points adaptively using each adaptive design. It can be observed that the computation time in seconds of the ARSD and the RCC methods are significantly less  than  that of ECL, EI, and LCB.
\begin{table}[!htp]
    \centering
\caption{\footnotesize  The average of the measurements $M_{C_{0}}$ in \nameref{ex3} for the adaptive designs ARSD, RCC, RCC-EI, ECL, EI, LCB and one-shot designs for $n_0=9$ and $N=\{27,36,45,54\}$ over 50 simulations for the contour level $a=\{4,5,6,6.6\}$. The values in the parentheses are the relative efficiency of the criteria over one-shot designs.}\label{Table_ex3}
\vspace{10pt}
\resizebox{15cm}{!}{
\setlength\tabcolsep{2pt}
  \begin{tabular}{r rrrrrrrrr}
    \hline 
  $N$~~~\textbf{one-shot} & \textbf{ARSD} &\textbf{{RCC}}& {\textbf{RCC-EI}}& \textbf{ECL} & \textbf{EI}& \textbf{LCB}&\\[0.05cm]
   \hline
 \hline& &&{a=4}&&&\\
   \hline
    27~~~~0.1909&0.1297(1.47)& 0.1587(1.20) &\textbf{0.1285(1.49)}& 0.1639(1.16)& 0.1289(1.48)&\textbf{0.1281(1.49)}\\    
    36~~~~0.1468&\textbf{0.0987(1.49)} & 0.1005(1.46)&\textbf{0.0982(1.49)}& 0.1068(1.37)& 0.1038(1.41)&0.1029(1.43)\\
    45~~~~0.1273&0.0719(1.77) & \textbf{0.0586(2.17)}&0.0811(1.57)& 0.0610(2.09)& 0.0867(1.47)&0.0745(1.71)\\
    54~~~~0.0922&0.0435(2.12)& \textbf{0.0269(3.43)} &0.0517(2.78)& 0.0315(2.93)& 0.0508(1.81)&0.0386(2.39)\\
    \hline& &&{a=5}&&&\\
   \hline
    27~~~~0.1572&0.1277(1.23) & 0.1492(1.05)&0.1298(1.21)& 0.1436(1.09)& \textbf{0.1219(1.29)}&0.1382(1.14)\\    
    36~~~~0.1118&\textbf{0.1032(1.08)}& 0.1090(1.03) &0.1089(1.03)& 0.1062(1.05)& 0.1075(1.04)&0.1029(1.09)\\
    45~~~~0.0917&{0.0930(0.98)}& \textbf{0.0832(1.10)} &0.0961(0.95)& \textbf{0.0828(1.10)}& 0.0951(0.96)&0.0917(1.00)\\
    54~~~~0.0634&0.0697(0.91)& \textbf{0.0486(1.30)} &0.0749(0.85)& 0.0503(1.26)& 0.0701(0.90)&0.0705(0.90)\\
    \hline& &&{a=6}&&&\\
   \hline
    27~~~~0.1450&0.1248(1.16) & 0.1338(1.08)&{0.1193(1.22)}& 0.1507(0.96)& \textbf{ 0.1167(1.24)}&{ 0.1177(1.23)}\\    
    36~~~~0.1200&0.0899(1.33) & 0.0930(1.29)&0.0929(1.29)& \textbf{0.0856(1.40)}& 0.0931(1.29)&0.0886(1.35)\\
    45~~~~0.1048&0.0582(1.80) & 0.0524(2.00)&0.0715(1.47)& \textbf{0.0498(2.10)}& 0.0720(1.46)&0.0623(1.66)\\
    54~~~~0.0744&0.0250(2.98)& \textbf{0.0218(3.41)} &0.0335(2.22)& 0.0220(3.36)& 0.0349(2.13)&0.0275(2.71)\\
 \hline& &&{a=6.6}&&&\\
 \hline

    27~~~~0.3554&\textbf{0.2866(1.24)} &{0.3200(1.11)}&0.3122(1.14)& \textbf{0.2870(1.24)}& 0.2872(1.24)&0.3232(1.10)\\   
    36~~~~0.2738&{0.1095(2.50)}& \textbf{0.0977(2.80)} &0.1701(1.61)& 0.1228(2.23)& 0.1599(1.71)&0.1777(1.54)\\
    45~~~~0.2413&0.0409(5.90) & \textbf{0.0278(8.86)}&0.0592(4.08)& 0.0308(7.83)& 0.0553(4.36)&0.0759(3.18)\\
    54~~~~0.1658&0.0221(7.50) & \textbf{0.0072(23.1)}&0.0110(15.1)& 0.0116(14.3)& 0.0122(13.5)&0.0361(4.59)\\
    \end{tabular}
    }
  \end{table}

  \begin{table}[!htp]
\footnotesize
   \centering
\caption{\footnotesize The average computation time per second in \nameref{ex3} for the adaptive designs ARSD, RCC, RCC-EI, ECL, EI, and LCB for $n_0=9$ and $N=54$ over 50 simulations for the contour level $a=\{4,5,6,6.6\}$.}\label{Ex3-time}
\vspace{10pt}
\resizebox{9cm}{!}{
\setlength\tabcolsep{2pt}
  \begin{tabular}{r rrrrrrrrrrrr}
    \hline 
  $a$~~~~~&&\textbf{ARSD}& &\textbf{{RCC}}& &\textbf{{RCC-EI}}& &\textbf{ECL} & &\textbf{EI}& &\textbf{LCB}\\[0.05cm]  \hline
   1.1~~~~~&&29.1& &31.6& &30.2& &86.7 & &85.1& &84.7\\
   1.5~~~~~&&59.2& &59.2& &60.1& &85.2 & &87.1& &88.3\\
   2~~~~~&&40.5& &39.8&&39.8& &85.5 & &87.5& &84.8\\
   2.6~~~~~&&23.3& &20.3& &21.0& &90.3 & &85.5& &85.5\\
      \end{tabular}
      }
  \end{table}

\section{Case Study of HPC data}\label{HPC} 

In this section, we apply the adaptive designs based on the RCC for studying high performance computing (HPC) systems, which are well known to be vital infrastructures to advance Industry 4.0.  
A key concept to evaluate the performance of the HPC systems is the HPC variability, run-to-run variation in the execution of a
computing task.  In particular, the input/output (IO)
throughput (i.e., data transfer speed) is an important
metric, which is affected by a number of system factors such as  the CPU frequency, the number of threads, the Input/Output (IO) file size, the IO record size, and the IO operation mode. Some of these system factors can be qualitative and quantitative. Hence the relation between the responses (IO throughput and variability) and these system factors can be quite complicated. A similar HPC application with different input variables and data has been considered in \cite{cai2024adaptive} where the goal of the study is to find the
level combination of system factors and the value of quantitative input variables that optimize
the IO performance variability measure.
Here we aim to find inputs such that the IO performance variability measure is close to certain values.

Table~\ref{Table-of-info} summarizes the input variables of which we have four quantitative variables,  the CPU clock frequency $({x}_1)$, the file size $({x}_2)$, the record size (${x}_3$) the number of thread $({x}_4)$, and one qualitative input factor the IO operation mode ${z}$ with six levels (initial writer, random readers, random writers, re-readers, readers and re-writers). For a given level combination of input factors as a configuration, the HPC server executes the \href{https://www.iozone.org/}{IOzone benchmark} task, and the IO throughput in kilobytes per second is recorded.  This process is repeated a large number of times \cite{cameron2019moana}. The mean and the standard deviation (SD) of the replicated IO throughput values are calculated. A smaller SD value indicates the HPC system's robustness. In contrast, a large mean indicates the effectiveness of the HPC system. Hence, 
we consider
to use the signal-to-noise ratio $Y$, i.e., the ratio of
the mean and SD of the throughput values, as the
output response.
A robust HPC system would require the value $Y$ to be larger than some threshold which is chosen by the domain expert. 
As the data shows right-skewness, we use the logarithm of $Y$  as the response variable in the model fitting in our study.  We show momentarily that our proposed method outperforms other competitors for multiple thresholds. For illustration purposes, we choose the contour levels to be  $a=\{24.5,27,30\}$.  

\begin{table}[!htp]
    \centering
\caption{  A summary of input variables in IO throughput experiment of the HPC system.}\label{Table-of-info}
\vspace{10pt}
\setlength\tabcolsep{2pt}
\begin{tabular}{p{2.5cm}p{4cm}p{3cm}p{5cm}}
    \hline 
    \textbf{Category} & \textbf{Variables} & \textbf{No. of levels} & \textbf{Values} \\[0.1cm]
    \hline  
    Hardware &${x}_1$: CPU Clock Frequency (GHz)& Continuous &1.2, 1.6, 2.0, 2.3, 2.8, 3.2, 3.5\\ \hline
    & ${x}_2$: File Size (KB) & Continuous &  4, 16, 64, 256,  1024,  4096,  8192, 16384, 32768, 65536\\ 
    Application & ${x}_3$: Record Size (KB) & Continuous &  4, 8, 16, 32, 64, 128, 256, 512, 1024, 2048,  4096,  8192, 16384 \\ 
    & ${x}_4$: Number of Threads & Continuous & 1, 8, 16, 24, 32, 40, 48, 56, 64 \\
    & ${z}$: IO Operation Mode & 6 & Initial writers, Random readers, Random writers, Re-readers, Readers, Rewriters \\ \hline
\end{tabular}
  \end{table}

We let $n_0=30$ and $N=150$.  We replicate the experiment 50 times.  For the candidate points in the design space, we consider 100 random LHDs for each level combination.  For the one-shot design, we consider 25 random LHDs for each level combination.
For measurement $M_{C_0}$, we consider $a\pm \epsilon$; as we are using logarithm transformation, a relatively small $\epsilon=0.006$ considered that the inputs are roughly the same if using $\epsilon=0.15$ for $Y$. This has been done to use the inputs that are close to actual values in $a=\{24.5,27,30\}$.  We set $\delta =0.5$ in Algorithm 3.1. Table \ref{Table-HPC} reveals that out of all adaptive designs, the relative efficiency of the proposed method is higher than other methods, although the efficiency gain is not as high as those in the simulation studies. Table \ref{HPC-time} shows the average computation time to add 120 points sequentially using each adaptive design. It can be observed that the computation time in seconds of the ARSD and the RCC methods are much less than those of other methods.

\begin{table}[!htp]
    \centering
\caption{\footnotesize  The average of the measurements $M_{C_{0}}$ in the HPC systems for the adaptive designs ARSD, RCC, RCC-EI, ECL, EI, LCB, and one-shot designs for $n_0=30$ and $N=\{135,140,145,150\}$ over 50 simulations for the contour level $a=\{24.5,27,30\}$. The values in the parentheses are the relative efficiency of the criteria over one-shot designs.}\label{Table-HPC}
\vspace{10pt}
\resizebox{15cm}{!}{
\setlength\tabcolsep{2pt}
  \begin{tabular}{r rrrrrrrrr}
    \hline 
  $N$~~~\textbf{one-shot} & \textbf{ARSD} &\textbf{{RCC}}& \textbf{{RCC-EI}}&\textbf{ECL} & \textbf{EI}& \textbf{LCB}&\\[0.05cm]
  \hline& &&{a=24.5}&&&\\
   \hline  
   135~~~~1.0860 & 0.9176(1.18)& 0.9418(1.15)& 0.9577(1.13)& 0.9657(1.12)& 0.9219(1.18)& \textbf{0.8821(1.23)}\\ 
  140~~~~1.1200 & \textbf{0.9018(1.24)}& 0.9093(1.23)& 0.9310(1.20)& 0.9828(1.14)& \textbf{0.8998(1.24)}& 0.9238(1.21)\\ 
   145~~~~1.1508 & 0.9436(1.22)& \textbf{0.8983(1.28)}& 0.9094(1.27)& 0.9629(1.20)& 0.9295(1.24)& 0.9178(1.25)\\ 
  150~~~~1.0890 & 0.9277(1.17)& \textbf{0.8848(1.23)}& 0.9554(1.14)& 0.9110(1.20)& 0.9134(1.19)& 0.8998(1.21)\\
  \hline& &&{a=27}&&&\\
   \hline
135~~~~0.8411 & 0.6901(1.22)& \textbf{0.6625(1.27)}& 0.6510(1.29)& 0.6749(1.25)& 0.7005(1.20)& 0.7108(1.18)\\ 
 140~~~~0.8374 & 0.6967(1.20)& \textbf{0.6521(1.28)}& 0.6833(1.23)& 0.6890(1.22)& 0.6852(1.22)& 0.7014(1.19)\\ 
  145~~~~0.8273 & 0.6900(1.20)& \textbf{0.6358(1.30)}& 0.6739(1.23)& 0.6720(1.23)& 0.6868(1.20)& 0.7058(1.17)\\ 
  150~~~~0.8587 & 0.6733(1.28)& \textbf{0.6447(1.33)}& 0.6799(1.26)& 0.6696(1.28)& 0.6751(1.27)& 0.6893(1.25)\\ 
 \hline& &&{a=30}&&&\\
  \hline
135~~~~0.8069 & 0.6671(1.21)& \textbf{0.6467(1.25)}& 0.6888(1.17)& 0.7176(1.12)& 0.6567(1.23)& 0.6706(1.20)\\ 
 140~~~~0.7925 & 0.6510(1.22)& 0.6549(1.21)& 0.6966(1.14)& 0.7315(1.08)& \textbf{0.6472(1.22)}& 0.6651(1.19)\\ 
 145~~~~0.8126 & \textbf{0.6362(1.28)}& 0.6462(1.26)& 0.6828(1.19)& 0.7340(1.11)& 0.6888(1.18)& 0.6738(1.21)\\ 
 150~~~~0.7964 & 0.6652(1.20)& \textbf{0.6179(1.29)}& 0.7112(1.12)& 0.7086(1.12)& 0.6363(1.25)& 0.6443(1.24)\\ 
\end{tabular}
}
  \end{table}

\begin{table}[!htp]
\footnotesize
   \centering
\caption{\footnotesize The average computation time (in seconds) in the HPC system for the adaptive designs ARSD, RCC, RCC-EI, ECL, EI, and LCB for $n_0=30$ and $N=150$ over 50 simulations for the contour level $a=\{24.5,27,30\}$. }\label{HPC-time}
\vspace{10pt}
\resizebox{9cm}{!}{
\setlength\tabcolsep{2pt}
  \begin{tabular}{r rrrrrrrrrrrr}
    \hline 
  $a$~~~~~&&\textbf{ARSD}& &\textbf{{RCC}}& &\textbf{{RCC-EI}}& &\textbf{ECL} & &\textbf{EI}& &\textbf{LCB}\\[0.05cm]  \hline
   24.5~~~~&&76.6& &97.7& &72.8& &117.2 & &120.5& &118.8\\
   27~~~~~&&72.3& &97.8& &71.4& &118.3 & &119.4& &121.5\\
   30~~~~~&&76.5& &92.9& &73.4& &122.1 & &124.9& &117.5\\
   \end{tabular}
   }
  \end{table}

 \section{Conclusion}
\label{sec:conc}
In this work, we propose an adaptive design approach for the contour estimation problem when computer experiments involve both quantitative and qualitative input variables. The proposed RCC method in the contour estimation problem divides the design space into two disjoint groups and considers different acquisition functions to choose the next point as discussed in Section \ref{sec:propos}. In this way, the advantages of different methods are combined and strengthened so that the proposed approach works well for different contour levels. 
Through the numerical comparison, we demonstrate in Section \ref{sec:simul} that our proposed method outperforms the one-shot designs, the EI, the ECL, the LCB, the RCC-EI, and the ARSD criteria across several contour levels. 
The theoretical justifications of the adaptive search region in one group are provided in Section \ref{sec:propos} for any contour level.

A few issues are worth highlighting. 
An effective adaptive design requires (a) a good surrogate model with the training data at hands; and (b) an efficient acquisition criterion to be solved to find the next input at which the computer model is run.  In this study, we apply the EzGP to build the surrogate although we also try the latent variable based Gaussian process (LvGP) models \cite{zhang2020latent} and similar results are obtained. However, both EzGP and LvGP are computationally expensive because of the number of parameters involved. Building a computationally efficient surrogate for computer experiments with mixed inputs remains an unsolved issue, particularly with the high dimensionality. 
Finding the mixed input that optimizes an acquisition criterion is an important yet challenging issue. In this article, we do not directly tackle this issue. Instead, we use a large number of candidate points to obtain a numerically approximate solution.  How to efficiently optimize the RCC criterion is worthy investigating and we leave it for future work. 
 
Another direction for future development is the contour estimation for large-scale computer experiments with both quantitative and qualitative inputs.  For the prediction at the new inputs for such experiments, \cite{xiao2021ezgp} proposed a method to address this issue   called localized EzGP. However, their model must select an appropriate tuning parameter for qualitative inputs to reduce the size of the data which still can be a large data when we have many level combinations. One future work we would like to investigate is addressing the contour estimation issue for large-scale computer experiments with quantitative and qualitative inputs by choosing informative local points for contour estimation.

\appendix
 \section{Appendix}\label{appendix}
 \subsection{Proof of Lemma \ref{lemma1}}\label{appendix_1}
For $\bm{w}_0\in \mathbb{A}$, consider $(Y(\bm{w}_0)-a)$ where $a$ is the contour of interest. We have,
\begin{align}\label{eq1lemma1}
  (Y(\bm{w}_0)-a) \sim N((\mu_{0|n}(\bm{w}_0)-a),\sigma_{0|n}^2(\bm{w}_0)),
\end{align}
where $\mu_{0|n}(\bm{w}_0)$ and $\sigma^2_{0|n}(\bm{w}_0)$ are the predicted mean and predicted variance. By standardizing (\ref{eq1lemma1}), we have, 

\begin{align*}
\begin{split}
    \frac{ (Y(\bm{w}_0)-a)-(\mu_{0|n}(\bm{w}_0)-a)}{\sigma_{0|n}(\bm{w}_0)}& \sim N(0,1).
\end{split}
\end{align*}
Now let $T \sim N(0,1)$ be a standard normal. Hence for $s>0$, we have, 
    \begin{align*}
   P(T>s)=&\frac{1}{\sqrt{2\pi}}\int_{s}^{\infty}\exp\{-\frac{1}{2}r^2\}dr=\frac{\exp\{-\frac{1}{2}s^2\}}{\sqrt{2\pi}}\int_{s}^{\infty}\exp\{-\frac{1}{2}(r-s)^2-s(r-s)\}dr.
\end{align*}
Then for $r\geq s>0$, $\exp\{-s(r-s)\}\leq 1$ and we have, 
\begin{align*}
    P(T>s)&\leq\frac{\exp\{-\frac{1}{2}s^2\}}{\sqrt{2\pi}}\int_{s}^{\infty}\exp\{-\frac{1}{2}(r-s)^2\}dr=\exp\{-\frac{1}{2}s^2\}P(T>0)=\frac{\exp\{-\frac{1}{2}s^2\}}{2}.
    \end{align*}
That is, $P(|T|>s)\leq \exp\{-\frac{1}{2}s^2\}$. Therefore, for $s=\sqrt{\beta_{0|n}}$, where $\beta_{0|n}=2\log(\frac{\pi^2n^2 M}{6\alpha})$ with $M=|\bm{Z}|=\prod_{j=1}^{q}m_j$ being the size of $\bm{Z}$ which is a finite discrete space and $\alpha\in(0,1)$ we have, 
\begin{align*}
    P(|Y(\bm{w}_0)-\mu_{0|n}(\bm{w}_0)|\geq \sqrt{\beta_{0|n}}\sigma_{0|n}(\bm{w}_0))\leq \exp\{-\frac{\beta_{0|n}}{2}\}.
\end{align*}
For a given input $\bm{x}_0\in \chi$ of the quantitative factors, by applying the union bound we get,

\begin{align*}
\begin{split}
    P(|Y(\bm{w}_0)-\mu_{0|n}(\bm{w}_0)|\leq\sqrt{\beta_{0|n}}\sigma_{0|n}(\bm{w}_0), \forall \bm{z}_0\in \bm{Z},\forall n\geq 1)\geq 1-M\sum_{n\geq 1}\exp\{\frac{-\beta_{0|n}}{2}\}.
\end{split}
\end{align*}
Now, consider $h(\bm{w}_0)=|Y(\bm{w}_0)-a)|$ and $\mu_h(\bm{w}_0)=|\mu_{0|n}(\bm{w}_0)-a|$. We wish to show, for a given input $\bm{x}_0\in \chi$ of the quantitative factors,
\begin{align}\label{proof1}
\begin{split}
    P(|h(\bm{w}_0)-\mu_{h}(\bm{w}_0)|\leq\sqrt{\beta_{0|n}}\sigma_{0|n}(\bm{w}_0),\forall \bm{z}_0 \in \bm{Z},\forall n\geq 1)> 1-\alpha.\\
\end{split}
\end{align}
To show this, we apply the reverse triangle inequality:
\begin{align*}
   ||b|-|c|| \leq |b-c|.
\end{align*}

Let $b=(Y(\bm{w}_0)-a)$, and $c=(\mu_{0|n}(\bm{w}_0)-a)$. We have, 

\begin{align*} 
\begin{split}
&P(|h(\bm{w}_0)-\mu_{h}(\bm{w}_0)|\leq\sqrt{\beta_{0|n}}\sigma_{0|n}(\bm{w}_0),\forall \bm{z}_0\in \bm{Z},\forall n\geq 1)\\
&\geq P(|(Y(\bm{w}_0)-a)-(\mu(\bm{w}_0)-a)|\leq \sqrt{\beta_{0|n}}\sigma(\bm{w}_0),\forall \bm{z}_0\in \bm{Z},\forall n\geq 1)\\
&\geq 1-M\sum_{n\geq 1}\exp\{\frac{-\beta_{0|n}}{2}\}\geq 1-M\sum_{n\geq 1}(\frac{6\alpha}{\pi^2n^2M})> 1-\alpha.\\
 \end{split}
\end{align*}
where $\alpha\in(0,1)$ and by the definition of $\beta_{0|n}$, (\ref{proof1}) holds.
\subsection{Proof of Lemma \ref{lemma2}}\label{appendix_2}
Recall from the definitions in (\ref{lemma2def}), 

\begin{align}
       &h_{\min}=\min_{\bm{w}_0\in \mathbb{A}}h(\bm{w}_0)=\sup_{\nu \in \mathbb{R}}\{\nu:\frac{1}{M}\sum_{\bm{z}_0\in \bm{Z}}\int_{\bm{x}_0\in \chi}\mathds{1}_{\{h(\bm{w}_0)<\nu\}}d{\bm{x}_0}<\epsilon\},\label{eq:subeq1}\\
       &\tilde{\mu}_{\min,n}=\min_{\bm{w}_0\in \mathbb{A}}\mu_h(\bm{w}_0)=\sup_{\nu\in \mathbb{R}}\{\nu:\frac{1}{M}\sum_{\bm{z}_0\in \bm{Z}}\int_{\bm{x}_0\in \chi}\mathds{1}_{\{\mu_h(\bm{w}_0)<\nu\}}d{\bm{x}_0}<\epsilon\}\label{eq:subeq2},\\
      & \tilde{\mu}^{L}_{\min,n}=\min_{\bm{w}_0\in\mathbb{A}}\mu^{L}_h(\bm{w}_0)=\sup_{\nu\in\mathbb{R}}\{\nu:\frac{1}{M}\sum_{\bm{z}_0\in \bm{Z}}\int_{\bm{x}_0\in \chi}\mathds{1}_{\{\mu^{L}_h(\bm{w}_0)<\nu\}}d{\bm{x}_0}<\epsilon\}\label{eq:subeq3},\\
   & \tilde{\mu}^{U}_{\min,n}=\min_{\bm{w}_0\in\mathbb{A}}\mu^{U}_h(\bm{w}_0)=\sup_{\nu\in\mathbb{R}}\{\nu:\frac{1}{M}\sum_{\bm{z}_0\in \bm{Z}}\int_{\bm{x}_0\in \chi}\mathds{1}_{\{\mu^{U}_h(\bm{w}_0)<\nu\}}d{\bm{x}_0}<\epsilon\}\label{eq:subeq4}.  
    \end{align}
where $\epsilon>0$ is a small positive number and $\mathds{1}(\cdot)$ is an indicator function. We will show Lemma \ref{lemma2} for any $a$ that is the contour of interest.  With the definition of $\mu_h^L(\bm{w}_0)$ and $\mu_h^U(\bm{w}_0)$ in (\ref{LBUB}), by Lemma \ref{lemma1},  we have,
\begin{align*}
    P(\mu^{L}_h(\bm{w}_0)\leq h(\bm{w}_0) \leq \mu^{U}_h(\bm{w}_0),\forall \bm{z}_0\in \bm{Z}, \forall n\geq 1 )> 1-\alpha.
\end{align*} 
When $\mu^{L}_{h}(\bm{w}_0)\leq h(\bm{w}_0) \leq \mu^{U}_{h}(\bm{w}_0)$, for $\nu \in \mathbb{R}$ and a given $\bm{x}_0\in \chi$, 
\begin{align*}
    \begin{split}
        \frac{1}{M}\sum_{\bm{z}_0\in \bm{Z}}\mathds{1}_{\{\mu^{U}_{h}(\bm{w}_0)<\nu\}}\leq  \frac{1}{M}\sum_{\bm{z}_0\in \bm{Z}}\mathds{1}_{\{h(\bm{w}_0)<\nu\}}\leq \frac{1}{M}\sum_{\bm{z}_0\in \bm{Z}}\mathds{1}_{\{\mu^{L}_{h}(\bm{w}_0)<\nu\}}.
    \end{split}
\end{align*}
For $\forall n\geq 1$, integrating over $\bm{x}_0$ we get, 
\begin{align}\label{26} 
    \begin{split}
        \int_{\bm{x}_0\in \chi}\frac{1}{M}\sum_{\bm{z}_0\in \bm{Z}}\mathds{1}_{\{\mu^{U}_{h}(\bm{w}_0)<\nu\}} d\bm{x}_0\leq & \int_{\bm{x}_0 \in \chi}\frac{1}{M}\sum_{\bm{z}_0\in \bm{Z}}\mathds{1}_{\{h(\bm{w}_0)<\nu\}}d\bm{x}_0      \\ 
        &\leq \int_{\bm{x}_0 \in \chi}\frac{1}{M}\sum_{\bm{z}_0\in \bm{Z}}\mathds{1}_{\{\mu^{L}_{h}(\bm{w}_0)<\nu\}}d\bm{x}_0. 
    \end{split}
\end{align}
Consider $\tilde{\mu}^{L}_{\min,n}$ is defined in (\ref{eq:subeq3}). Let $\nu=\tilde{\mu}^{L}_{\min,n}$ in (\ref{26}) we get that, 
\begin{align}
&\frac{1}{M}\sum_{\bm{z}_0\in \bm{Z}}\int_{\bm{x}_0\in \chi}\mathds{1}_{\{h(\bm{w}_0)<\tilde{\mu}^{L}_{\min,n}\}}d{\bm{x}_0}=\int_{\bm{x}_0\in \chi}\frac{1}{M}\sum_{\bm{z}_0\in \bm{Z}}\mathds{1}_{\{h(\bm{w}_0)<\tilde{\mu}^{L}_{\min,n}\}}d{\bm{x}_0} \label{Cm1}\\
\leq &\int_{\bm{x}_0\in \chi}\frac{1}{M}\sum_{\bm{z}_0\in \bm{Z}}\mathds{1}_{\{\mu^{L}_{h}(\bm{w}_0)<\tilde{\mu}^{L}_{\min,n}\}}d{\bm{x}_0}=\frac{1}{M}\sum_{\bm{z}_0\in \bm{Z}}\int_{\bm{x}_0\in \chi}\mathds{1}_{\{\mu^{L}_{h}(\bm{w}_0)<\tilde{\mu}^{L}_{\min,n}\}}d{\bm{x}_0}<\epsilon \nonumber
\end{align}
As $h_{\min}=\sup\{\frac{1}{M}\sum_{\bm{z}_0\in \bm{Z}}\int_{\bm{x}_0\in \chi}\mathds{1}_{\{h(\bm{w}_0)<\tilde{\mu}^{L}_{\min,n}\}}d{\bm{x}_0}<\epsilon\}$, (\ref{Cm1}) reveals that $\tilde{\mu}^{L}_{\min,n} \leq h_{\min}$. Let $\nu=h_{\min}$ in (\ref{26}) and by the definition of $h_{\min}$ in (\ref{eq:subeq1}) we get, 
\begin{align}
&\frac{1}{M}\sum_{\bm{z}_0\in \bm{Z}}\int_{\bm{x}_0\in \chi}\mathds{1}_{\{\mu^{U}_h(\bm{w}_0)<h_{\min}\}}d{\bm{x}_0}=\int_{\bm{x}_0\in \chi}\frac{1}{M}\sum_{\bm{z}_0\in \bm{Z}}\mathds{1}_{\{\mu^{U}_h(\bm{w}_0)<h_{\min}\}}d{\bm{x}_0} \label{26+2} \\\leq &\int_{\bm{x}_0 \in \chi}\frac{1}{M}\sum_{\bm{z}_0\in \bm{Z}}\mathds{1}_{\{h(\bm{w}_0)<h_{\min}\}}d\bm{x}_0= \frac{1}{M}\sum_{\bm{z}_0\in \bm{Z}}\int_{\bm{x}_0 \in \chi}\mathds{1}_{\{h(\bm{w}_0)<h_{\min}\}}d\bm{x}_0<\epsilon. \nonumber
\end{align}
As $\tilde{\mu}^{U}_{\min,n}=\sup\{\frac{1}{M}\sum_{\bm{z}_0\in \bm{Z}}\int_{\bm{x}_0\in \chi}\mathds{1}_{\{\mu^{U}_h(\bm{w}_0)<h_{\min}\}}d{\bm{x}_0}<\epsilon\}$, (\ref{26+2}) indicate that $h_{\min} \leq \tilde{\mu}^{U}_{\min,n}$. Therefore we have, $\tilde{\mu}^{L}_{\min,n} \leq h_{\min}\leq \tilde{\mu}^{U}_{\min,n}$ when (\ref{26}) holds. As (\ref{26}) occurs with the probability greater than $1-\alpha$, 
\begin{align*}
    \begin{split}
P([\tilde{\mu}^{L}_{\min,n}\leq h_{\min} \leq \tilde{\mu}^{U}_{\min,n}],\forall n\geq 1)>1-\alpha.
    \end{split}
\end{align*}
This concludes the proof of Lemma \ref{lemma2}.
\subsection{Proof of Theorem \ref{Theorem1}}\label{appendix_4}
To prove Theorem \ref{Theorem1}, we need to show for any input in the search space $\mathbb{A}_c=\mathbb{A}_{1,\min}\cup \mathbb{A}_2$, we have,
\begin{align}\label{claim1}
        \tilde{\mu}^{U}_{\min,n}\leq \tilde{\mu}_{{\min,n}}+\sqrt{\beta_{0|n}}\sup_{\bm{w}_0\in\mathbb{A}_c}\sigma_{0|n}(\bm{w}_0).
\end{align}
By (\ref{eq:subeq4}), for all $n\geq1$,
    $\frac{1}{M}\sum_{\bm{z}_0\in \bm{Z}}\int_{\bm{x}_0\in \chi}\mathds{1}_{\{\mu^{U}_{h}(\bm{w}_0)<\tilde{\mu}^{U}_{\min,n}\}}d{\bm{x}_0}<\epsilon.$ For a given $\bm{z}_0$, define
\begin{align}\label{An0}
    \mathbb{A}_{\chi}(\bm{z}_0)=\{\bm{x}_0\in \chi:\mu_h^{L}(\bm{x}_0,\bm{z}_0)\leq 
    \tilde{\mu}^{U}_{\min,n}\}.
\end{align}
It is clear that, $\mathbb{A}_c=\{(\bm{x}_0,\bm{z}_0):\bm{z}_0\in \bm{Z}, \bm{x}_0\in \mathbb{A}_{\chi}(\bm{z}_0)\}$.  Because if $\bm{x}_0\notin\mathbb{A}_{\chi}(\bm{z}_0)$, that is, $\tilde{\mu}^{U}_{\min,n}<\mu_h^{L}(\bm{x}_0,\bm{z}_0)$. Therefore $\tilde{\mu}^{U}_{\min,n}<\mu_h^{U}(\bm{x}_0,\bm{z}_0)$. Thus $\int_{\bm{x}_0\notin \mathbb{A}_{\chi}(\bm{z}_0)}\mathds{1}_{\{\mu^{U}_{h}(\bm{w}_0)< \tilde{\mu}^{U}_{\min,n}\}}d{\bm{x}_0}=0$. That is, we have,
\begin{align*}
    \begin{split}
        \frac{1}{M}\sum_{\bm{z}_0\in \bm{Z}}\int_{\bm{x}_0\in \mathbb{A}_{\chi}(\bm{z}_0)}\mathds{1}_{\{\mu^{U}_{h}(\bm{w}_0)< \tilde{\mu}^{U}_{\min,n}\}}d{\bm{x}_0}&=\frac{1}{M}\sum_{\bm{z}_0\in \bm{Z}}\int_{\bm{x}_0\in \chi}\mathds{1}_{\{\mu^{U}_{h}(\bm{w}_0)< \tilde{\mu}^{U}_{\min,n}\}}d{\bm{x}_0}<\epsilon.
    \end{split}
\end{align*}
From the definition of $\mu^{U}_h(\bm{w}_0)$ given in (\ref{LBUB}) we obtain, 

\begin{align*}
    \begin{split}
        &~~~~\frac{1}{M}\sum_{\bm{z}_0\in \bm{Z}}\int_{\bm{x}_0\in\mathbb{A}_{\chi}(\bm{z}_0)} \mathds{1}_{\{\mu_{h}(\bm{w}_0)<\tilde{\mu}^{U}_{\min,n}-\sqrt{\beta_{0|n}}\sup_{\bm{w}_0\in\mathbb{A}_c}\sigma_{0|n}(\bm{w}_0)\}}d{\bm{x}_0}\\
        &\leq \frac{1}{M}\sum_{\bm{z}_0\in \bm{Z}}\int_{\bm{x}_0\in\mathbb{A}_{\chi}(\bm{z}_0)}\mathds{1}_{\{\mu_h(\bm{w}_0)<\tilde{\mu}^{U}_{\min,n}-\sqrt{\beta_{0|n}}\sigma_{0|n}(\bm{w}_0)\}}d\bm{x}_0<\epsilon.
    \end{split}
\end{align*}
Therefore by the definition  of $\mathbb{A}_{\chi}(\bm{z}_0)$ in (\ref{An0}) we get,
\begin{align*}
    \begin{split}
        &~~~~\frac{1}{M}\sum_{\bm{z}_0\in \bm{Z}}\int_{\bm{x}_0\in \chi}\mathds{1}_{\{\mu_{h}(\bm{w}_0)<\tilde{\mu}^{U}_{\min,n}-\sqrt{\beta_{0|n}}\sup_{\bm{w}_0\in\mathbb{A}_c}\sigma_{0|n}(\bm{w}_0)\}}d{\bm{x}_0}\\
        &=\frac{1}{M}\sum_{\bm{z}_0\in \bm{Z}}\int_{\bm{x}_0\in\mathbb{A}_{\chi}(\bm{z}_0)}\mathds{1}_{\{\mu_{h}(\bm{w}_0)<\tilde{\mu}^{U}_{\min,n}-\sqrt{\beta_{0|n}}\sup_{\bm{w}_0\in\mathbb{A}_c}\sigma_{0|n}(\bm{w}_0)\}}d\bm{x}_0<\epsilon.
    \end{split}
\end{align*}
Thus by (\ref{eq:subeq2}), the definition of $\tilde{\mu}_{\min,n}$, considering that $\epsilon>0$ and a small number we have, $\tilde{\mu}_{\min,n}\geq \tilde{\mu}^{U}_{\min,n}-\sqrt{\beta_{0|n}}\sup_{\bm{w}_0\in\mathbb{A}_c}\sigma_{0|n}(\bm{w}_0).$ That is,

\begin{align*}
    \tilde{\mu}^{U}_{{\min,n}}\leq \tilde{\mu}_{\min,n}+\sqrt{\beta_{0|n}}\sup_{\bm{w}_0\in\mathbb{A}_c}\sigma_{0|n}(\bm{w}_0).
\end{align*}
Therefore we obtained (\ref{claim1}), therefore, 
\begin{align*}
    h_{\min}\leq\tilde{\mu}^{U}_{{\min,n}}\leq \tilde{\mu}_{\min,n}+\sqrt{\beta_{0|n}}\sup_{\bm{w}_0\in\mathbb{A}_c}\sigma_{0|n}(\bm{w}_0).
\end{align*}
On the other hand, we need to show that, 
\begin{align}\label{claim2}
       \tilde{\mu}_{\min,n} \leq \tilde{\mu}^{L}_{\min,n}+\sqrt{\beta_{0|n}}\sup_{\bm{w}_0\in\mathbb{A}_c}\sigma_{0|n}(\bm{w}_0).
\end{align}
Note that, for all $n \ge 1$, ${\mu}^{L}_{h}(\bm{w}_0)\leq \mu_{h}(\bm{w}_0) \leq {\mu}^{U}_{h}(\bm{w}_0)$, by using the same argument we get that, $\tilde{\mu}^{L}_{\min,n}\leq \tilde{\mu}_{\min,n} \leq \tilde{\mu}^{U}_{\min,n}$ for all $n\geq1$.
By (\ref{eq:subeq2}) for any $n\geq1$ we have,

\begin{align*}
    \frac{1}{M}\sum_{\bm{z}_0\in \bm{Z}}\int_{\bm{x}_0\in \chi}\mathds{1}_{\{\mu_{h}(\bm{w}_0)<\tilde{\mu}_{\min,n}\}}d{\bm{x}_0}<\epsilon.
\end{align*}
Because if $\bm{x}_0\notin\mathbb{A}_{\chi}(\bm{z}_0)$, ${\mu}^{L}_{h}(\bm{x}_0,\bm{z}_0)>\tilde{\mu}^{U}_{\min,n}$, and  we have $\int_{\bm{x}_0\notin \mathbb{A}_{\chi}(\bm{z}_0)}\mathds{1}_{\{\mu_{h}(\bm{w}_0)< \tilde{\mu}_{\min,n}\}}d{\bm{x}_0}=0$. Therefore,
\begin{align*}
    \begin{split}
        \frac{1}{M}\sum_{\bm{z}_0\in \bm{Z}}\int_{\bm{x}_0\in \mathbb{A}_{\chi}(\bm{z}_0)}\mathds{1}_{\{\mu_{h}(\bm{w}_0)< \tilde{\mu}_{\min,n}\}}d{\bm{x}_0}&=\frac{1}{M}\sum_{\bm{z}_0\in \bm{Z}}\int_{\bm{x}_0\in \chi}
        \mathds{1}_{\{\mu_{h}(\bm{w}_0)< \tilde{\mu}_{\min,n}\}}d{\bm{x}_0}<\epsilon.
    \end{split}
\end{align*}
From the definition of $\mu_{0|n}^{L}(\bm{w}_0)$ in (\ref{LBUB}), and ${\mu}^{L}_{h}(\bm{w}_0)\leq \mu_{h}(\bm{w}_0) \leq {\mu}^{U}_{h}(\bm{w}_0)$ and $\tilde{\mu}^{L}_{\min,n}-\sqrt{\beta_{0|n}}\sup_{\bm{w}_0\in\mathbb{A}_c}\sigma_{0|n}(\bm{w}_0)\leq \tilde{\mu}_{\min,n}-\sqrt{\beta_{0|n}}\sup_{\bm{w}_0\in\mathbb{A}_c}\sigma_{0|n}(\bm{w}_0)$, then we have,
\begin{align*}
    \begin{split}
        &~~~~\frac{1}{M}\sum_{\bm{z}_0\in \bm{Z}}\int_{\bm{x}_0\in\mathbb{A}_{\chi}(\bm{z}_0)}\mathds{1}_{\{\mu^{L}_h(\bm{w}_0)<\tilde{\mu}_{\min,n}-\sqrt{\beta_{0|n}}\sup_{\bm{w}_0\in\mathbb{A}_c}\sigma_{0|n}(\bm{w}_0)\}}d{\bm{x}_0}\\
        &=\frac{1}{M}\sum_{\bm{z}_0\in \bm{Z}}\int_{\bm{x}_0\in \mathbb{A}_{\chi}(\bm{z}_0)}\mathds{1}_{\{\mu_h(\bm{w}_0)-\sqrt{\beta_{0|n}}\sigma_{0|n}(\bm{w}_0)<\tilde{\mu}_{\min,n}-\sqrt{\beta_{0|n}}\sup_{\bm{w}_0\in\mathbb{A}_c}\sigma_{0|n}(\bm{w}_0)\}}d\bm{x}_0\\
        &\leq \frac{1}{M}\sum_{\bm{z}_0\in \bm{Z}}\int_{\bm{x}_0\in\mathbb{A}_{\chi}(\bm{z}_0)}\mathds{1}_{\{\mu_h(\bm{w}_0)-\sqrt{\beta_{0|n}}\sup_{\bm{w}_0\in\mathbb{A}_
    c}\sigma_{0|n}(\bm{w}_0)<\tilde{\mu}_{\min,n}-\sqrt{\beta_{0|n}}\sup_{\bm{w}_0\in\mathbb{A}_c}\sigma_{0|n}(\bm{w}_0)\}}d\bm{x}_0<\epsilon.
    \end{split}
\end{align*}
Therefore by the definition  of $\mathbb{A}_{\chi}(\bm{z}_0)$ in (\ref{An0}) and $\tilde{\mu}^{L}_{\min,n}\leq \tilde{\mu}_{\min,n} \leq \tilde{\mu}^{U}_{\min,n}$,
\begin{align*}
    \begin{split}
      &\frac{1}{M}\sum_{\bm{z}_0\in \bm{Z}}\int_{\bm{x}_0\in\mathbb{A}_{\chi}(\bm{z}_0)}\mathds{1}_{\{\mu^{L}_h(\bm{w}_0)<\tilde{\mu}_{\min,n}-\sqrt{\beta_{0|n}}\sup_{\bm{w}_0\in\mathbb{A}_c}\sigma_{0|n}(\bm{w}_0)\}}d{\bm{x}_0}\\
      &= \frac{1}{M}\sum_{\bm{z}_0\in \bm{Z}}\int_{\bm{x}_0\in \chi}\mathds{1}_{\{\mu^{L}_h(\bm{w}_0)<\tilde{\mu}_{\min,n}-\sqrt{\beta_{0|n}}\sup_{\bm{w}_0\in\mathbb{A}_c}\sigma_{0|n}(\bm{w}_0)\}}d{\bm{x}_0}<\epsilon.
    \end{split}
\end{align*}
Thus, by the definition of $\tilde{\mu}^{L}_{\min,n}$ and considering a small number $\epsilon>0$  we have  $\tilde{\mu}^{L}_{\min,n} \geq \tilde{\mu}_{\min,n}-\sqrt{\beta_{0|n}}\sup_{\bm{w}_0\in\mathbb{A}_c}\sigma_{0|n}(\bm{w}_0)$, that is, 
\begin{align*}
   \tilde{\mu}_{\min,n} \leq\tilde{\mu}^{L}_{\min,n}+\sqrt{\beta_{0|n}}\sup_{\bm{w}_0\in\mathbb{A}_c}\sigma_{0|n}(\bm{w}_0).
\end{align*}
We obtained (\ref{claim2}) and by Lemma \ref{lemma2}, we have $P(\tilde{\mu}^{L}_{\min,n} \leq h_{\min} \leq \tilde{\mu}^{U}_{\min,n},\forall n\geq 1)> 1-\alpha$. Therefore, 
\begin{align*}
   \tilde{\mu}_{\min,n} \leq \tilde{\mu}^{L}_{\min,n}+\sqrt{\beta_{0|n}}\sup_{\bm{w}_0\in\mathbb{A}_c}\sigma_{0|n}(\bm{w}_0)\leq h_{\min}+\sqrt{\beta_{0|n}}\sup_{\bm{w}_0\in\mathbb{A}_c}\sigma_{0|n}(\bm{w}_0).
\end{align*}
From result of Lemma \ref{lemma1} and Lemma \ref{lemma2}, and we showed (\ref{claim1}), (\ref{claim2}) therefore we have,
\begin{align*}
    P(|\tilde{\mu}_{\min,n}-h_{\min}|\leq \sqrt{\beta_{0|n}}\sup_{\bm{w}_0\in\mathbb{A}_c}\sigma_{0|n}(\bm{w}_0)\forall n\geq 1)> 1-\alpha.
\end{align*}
Therefore Theorem \ref{Theorem1} holds for $\alpha\in (0,1)$.

\section*{Acknowledgments}
{Lin and Shahrokhian were supported by Discovery grant from the Natural Sciences and Engineering
Research Council of Canada. }

\bibliography{references}

\end{document}